\documentclass[iop]{emulateapj}
\usepackage{color}
\usepackage{graphicx}
\usepackage{xcolor}

\usepackage[]{natbib}
 
\newcommand{\simgt}{\lower.5ex\hbox{$\; \buildrel > \over \sim \;$}}
\newcommand{\simlt}{\lower.5ex\hbox{$\; \buildrel < \over \sim \;$}}

\newcommand{\Mvir}{M_\mathrm{vir}}
\newcommand{\Mhalo}{M_\mathrm{200c}}
\newcommand{\cvir}{c_\mathrm{vir}}
\newcommand{\chalo}{c_\mathrm{200c}}

\newcommand{\rc}{r_\mathrm{c}}
\newcommand{\rs}{r_\mathrm{s}}
\newcommand{\rhalo}{r_\mathrm{200c}}

\def\btheta{\mbox{\boldmath $\theta$}}

\newcommand{\etal}{et al.~}
\newcommand{\eg}{e.g.~}
\newcommand{\XMM}{XMM-{\em Newton}~}
\newcommand{\Chandra}{{\em Chandra}~}
\newcommand{\kms}{km\,s$^{-1}$}
\def\degree{\mbox{$^{\circ}$}}
\newcommand{\etc}{etc.~}

\begin{document}

\shorttitle{THE DOUBLE GALAXY CLUSTER ABELL 2465 III.}
\shortauthors{G. A. Wegner \etal}

\title{The double galaxy cluster Abell 2465 III. X-ray and weak-lensing observations\altaffilmark{*}}
\author{Gary A. Wegner\altaffilmark{1}} 
\author{Keiichi Umetsu\altaffilmark{2}} 
\author{Sandor M. Molnar\altaffilmark{2}}
\author{Mario Nonino\altaffilmark{3}} 
\author{Elinor Medezinski\altaffilmark{4}}
\author{Felipe Andrade-Santos\altaffilmark{5}}
\author{Akos Bogdan\altaffilmark{5}}
\author{Lorenzo Lovisari\altaffilmark{5}}
\author{William R. Forman\altaffilmark{5}}
\author{Christine Jones\altaffilmark{5}}

\altaffiltext{1}{Department of Physics \& Astronomy, 6127 Wilder Laboratory, 
 Hanover, NH 03745, USA}
 \email{gary.wegner@dartmouth.edu}
\altaffiltext{2}{Institute of Astronomy and Astrophysics, Academia Sinica, P. O. 
Box 23-141, Taipei 10617, Taiwan}
\altaffiltext{3}{INAF/Osservatorio Astronomico di Trieste, via G. B. Tiepolo 11,
34143 Trieste, Italy}
\altaffiltext{4}{Department of Astrophysical Sciences, 4 Ivy Lane, Princeton,
NJ 08544, USA}
\altaffiltext{5}{Harvard-Smithsonian Center for Astrophysics, 60 Garden Street, 
Cambridge, MA 02138, USA}

\altaffiltext{*}{Based in part on data collected at the Subaru Telescope,
which is operated by the National Astronomical Society of Japan.}

\begin{abstract}
We report \Chandra X-ray observations and optical weak-lensing
 measurements from Subaru/Suprime-Cam images of the double galaxy
 cluster  
 Abell 2465 ($z = 0.245$).
 The X-ray brightness data are fit to a
 $\beta$-model to obtain the radial gas density profiles of the
 northeast (NE) and southwest (SW) sub-components, which are seen to
 differ in structure.
We determine core radii, central temperatures, the gas masses within
 $r_\mathrm{500c}$, and the total masses for the broader NE and sharper
 SW components assuming hydrostatic equilibrium.
 The central entropy of the NE clump is about two times
 higher than the SW.
 Along with its structural properties, this suggests that it has
 undergone merging on its own.
The weak-lensing analysis gives virial masses for each 
substructure, which compare well with earlier dynamical results.
The derived outer mass contours of the SW sub-component from weak
 lensing are more irregular and extended than those of the NE.  
Although there is a weak enhancement and small offsets between
X-ray gas and mass centers from weak lensing, 
the lack of large amounts of gas between the two sub-clusters
indicates that Abell 2465 is in a pre-merger state. 
A dynamical model that is consistent with the observed cluster data,
based on the FLASH program and the radial infall model, 
 is constructed,
 where the subclusters currently separated by $\sim 1.2$\,Mpc are approaching each 
other at $\sim 2000$\,\kms and will meet in $\sim 0.4$\,Gyr.  
\end{abstract}


\section{Introduction}

In the $\Lambda$ cold dark matter ($\Lambda$CDM) picture of large-scale
structure formation, galaxy clusters grow hierarchically from smaller
knots of higher density forming and merging along the intersection of filaments.
Examples of interacting clusters that have undergone merging 
interactions include \eg the 'Bullet' (Clowe \etal 2006), RXJ1347.5-1145,
(Brada\v{c} \etal 2008a), MACSJ0025.4-1222 (Brada\v{c} \etal 2008b), 
Abell 2744 (Owers \etal 2011), Abell 2146 (Russell \etal 2012), 
and DLSCL J00916.2+2915 (Dawson \etal 2012). 
Some examples of pre-collisional binary clusters include
Abell 3716 (Andrade-Santos \etal 2015) and
Abell 1750 and 1758 (Okabe \& Umetsu 2008). Other possible objects have
been given in Hwang \& Lee (2009)and Molnar \etal (2013)
In X-rays, the 
structures of the gas in
galaxy clusters can be compared with the other components and this indicates
a range of structures extending from circular symmetry, indicating relaxed 
objects, to separated and disturbed gas indicating core-crossing 
events ranging from pre- to post-mergers (\eg Markevitch \&
Vikhlinin 2007). 

Modelling cluster collisions indicates the different behavior of
the collisionless dark matter and baryonic gas constituents 
(\eg Roettiger \etal 1997; Takizawa 2000; Ricker \& Sarazin
2001; Poole \etal 2006; Molnar, Hearn, \& Stadel 2012). Several 
investigators (\eg Spergel \& Steinhardt 2000; Kahlhoefer \etal 2014) 
have proposed utilizing these effects to study nongravitational 
interactions of dark matter 
in the collisions of the galaxy clusters to gain information on the 
properies of the dark matter component.
Vijayaragharan \& Ricker (2013) have shown that these effects begin to be 
felt even in the pre-merger phases so details of the dynamics
at all stages of cluster mergers provide rich information about the physical
interplay between dark matter and baryons.
 
Investigating double galaxy clusters
by combining optical, radio, and X-ray data with gravitational lensing provides
insight into the star formation and the evolution of of these
clusters and ultimately large-scale structure. 
Weak lensing is a powerful tool for reconstructing cluster mass distributions
on large angular scales and for identifying mass substructures 
(Okabe \& Umetsu 2008; Medezinski \etal 2013; Umetsu \etal 2012).
In many cases, the substructure of galaxy clusters complicates their 
interpretation if they have many components (\eg Cohen \etal 2014;
Merten \etal 2011; Medezinski \etal 2016) and consequently 
finding simple clearly double structures is valuable for elucidating the
dynamics of their components.

Abell 2465 appears to be a well defined double cluster system 
undergoing a major merger. 
First reported by Abell (1958) as a modest Richness Class 1 cluster, 
its double nature
was not known until Vikhlinin \etal (1998) gave ROSAT X-ray fluxes of the
two sub-components, referred to in this paper as the northeast (NE) and 
southwest (SW) components (See Figure~\ref{isodensity}).
Perlman \etal (2002) found redshifts of $z = 0.245$ for both, establishing
their physical relationship. 
The NE component of the cluster is detected in the
FIRST 1.4\,Ghz survey (Condon \etal 1998; Helfand, White, \& Becker 2015).

The optical properties of Abell 2465 have been described in Wegner
(2011; Paper I) and Wegner, Chu, \& Hwang (2015; Paper II). Virial
masses  
$\Mvir = (4.1 \pm 0.8) \times 10^{14}M_\odot$ and 
$(3.8 \pm 0.8) \times 10^{14} M_\odot$
(all relevant symbols are defined at the end of this section)
for NE and SW respectively were derived from optical velocity dispersions. 
Although they have similar masses, the two
subclusters differ in their radial profiles. 
The NE is less compact, while the SW is smaller in 
extent with a brighter inner core. In Paper II, it was
found that star formation rates of member galaxies appear enhanced.
Since the projected separation between the two subclusters is 1.2\,Mpc
and their optical halo radii are $\rhalo \approx 1.2$ and $1.25$\,Mpc
(Paper I), detectable effects of their interaction seem possible.

To look for such effects, we obtain and analyze
new {\it Chandra} X-ray observations
and weak-lensing measurements from Subaru Suprime-Cam imaging. 
These are utilized to determine the state of the baryonic gas 
and measure the structures and mass distibutions
of the double components of Abell 2465, study
their interaction, and settle whether they are pre- or post-core crossing. 

This paper is arranged as follows: Section 2 discusses the X-ray data.
Section 2.1 presents the {\it Chandra}
observations and reductions. Section 2.2 analyzes the X-ray
spatial and spectral data and the resulting total and gas masses
for the NE and SW subclusters. Section 2.3 determines the gas
temperaures of the substructures and Section 2.4 
derives the individual entropies. Section 2.5 describes the search for
gas between the subclusters. Section 2.6 further presents the optical
appearances of the components. 
Section 3  contains the optical imaging data and weak-lensing results for
Abell 2465. Section 3.1 covers the observations, Sections 3.2 and 3.3
explain the shape measurements and background selection
for the weak-lensing analysis. Multi-halo mass modeling is in Section 3.4.   
Section 4 concerns the dynamical state of and modeling of
Abell 2465. Section 5 discusses these results and how the cluster compares
with other merging galaxy clusters. Section 6 lists our conclusions.

We assume a standard $\Lambda$CDM cosmology with
$\Omega_\mathrm{m} = 0.3$, 
$\Omega_\Lambda = 0.7$, and
$H_0 = 100h$\,km\,s$^{-1}$\,Mpc$^{-1}$
with $h=0.7$.
For the cluster's mean redshift, $z = 0.245$, the luminosity distance is 
$D_L =1224$\,Mpc and the scale on the sky is 230\,kpc arcmin$^{-1}$. The 
two subclusters are separated by 5.25 arcmin or 1.2\,Mpc.

We use the standard notation $M_{\Delta \mathrm{c}}$ to denote the mass  
enclosed within a sphere of radius $r_{\Delta\mathrm{c}}$, within which
the mean  overdensity is $\Delta_\mathrm{c}$ times the critical density
$\rho_\mathrm{c}$ of the universe at the cluster redshift. To calculate
halo virial  quantities, we use an expression for the virial overdensity
$\Delta_\mathrm{vir}$ based on the spherical collapse model (see
Appendix A of Kitayama \& Suto 1996).

\section{Properties if Abell 2465 from X-ray observations}
\subsection{\Chandra X-ray Data}

Abell 2465 was observed 2012 October 6 and 2 in the 0.1-10 keV energy
range with the ACIS-I detectors of
{\it Chandra}, ObsIds 14010 and 15547 for 40\,ks and 30\,ks respectively.
Both the NE and SW sub-components were well placed together across the
instrument. 
With \Chandra, about 1800 net counts were collected from each 
sub-cluster after background subtraction. Data reductions applied 
CALDB 4.6.7 and followed Vikhlinin
\etal (2005). The reductions used the Chandra Interactive Analysis
of Observations (CIAO) package and included
calibration corrections to the individual photons,
calibration of the spectral response,
background subtraction including
quiescent and soft background correction, and
subtraction of the readout artifacts.
We reprocessed the data, excluded flare contaminated time intervals, 
detected and excluded point sources, merged the data sets, utilized ACIS
blank sky observations to subtract the background, and removed
readout artifacts.

Although \XMM images of Abell 2465 are available (Paper I), 
we only utilize the newer \Chandra observations. Both data sets have
similar number counts, but the cluster falls at the edge of the 
serendipitous \XMM
field of view which badly degrades the point spread function
and renders these data unsuitable for point source removal and producing
surface brightness, emission measure, and gas mass profiles.

\begin{figure*}
\epsscale{1.0}
\plotone{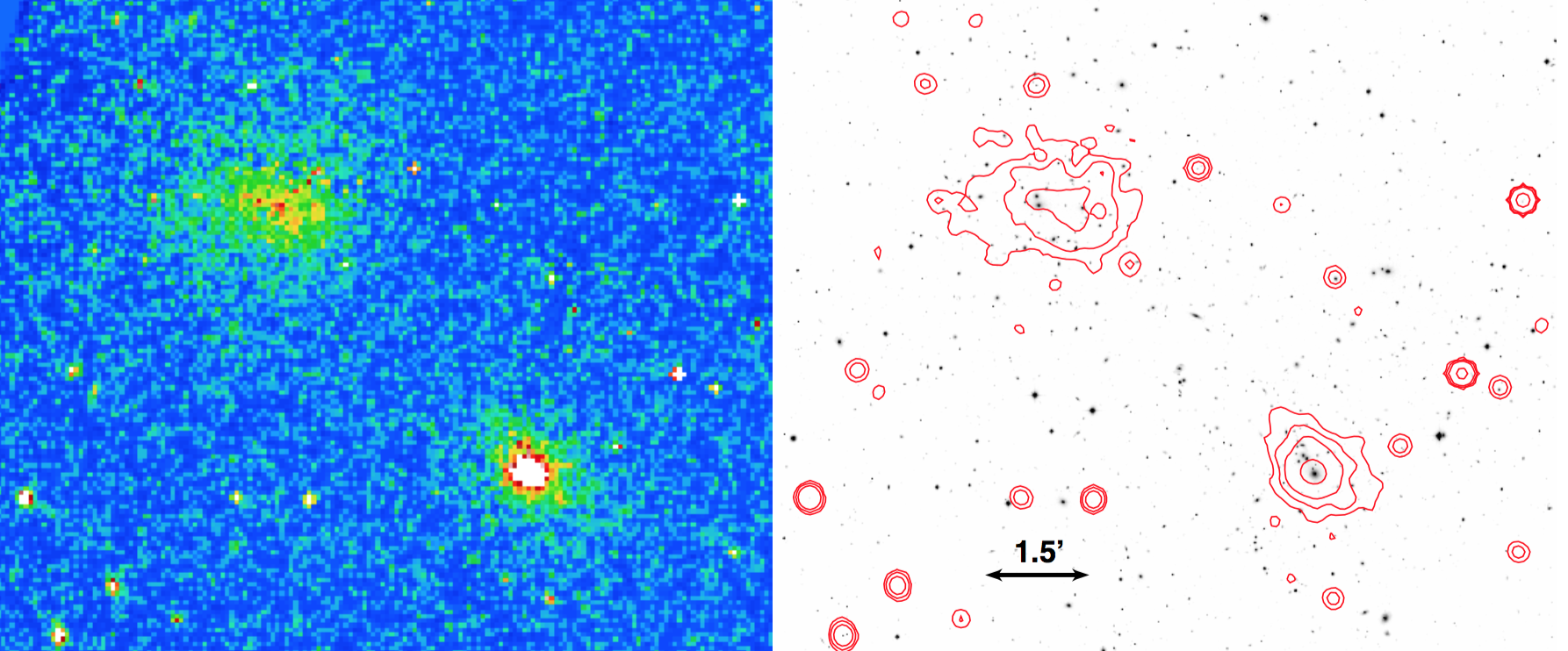}
\caption{Left: \Chandra image of Abell 2465 in the 0.5 to 3.0\,keV
 energy band with 8 pixel binning.  
Right: CFHT megaprime $i'$ image with the isodensity X-ray
 contours. North is to the top and West to the right. The images are
 approximately 
 $11 \times 9$\,arcmin$^2$.
The NE component is to the upper left and comparison with the SW
 component shows the differences in their structures. The X-ray center
 of the NE sub-cluster is slightly displaced from its central BCGs while
 the SW X-ray peak is nearly centered on the BCG. 
}
\label{isodensity}
\end{figure*}

Figure~\ref{isodensity} presents both
the resulting \Chandra image and its isodensity 
contours overlaying the optical $i'$ image from the 3.6\,m
Canada--France--Hawaii Telescope (CFHT; see Section \ref{subsec:subaru}).
This shows the different appearances of the two sub-clusters.
The X-ray profile of the NE component, although it is the more massive
and luminous, has a broad central maximum. It has two X-ray 
concentrations oriented E and W. The
SW peak is the brighter and is not near a radio or optical source. The
NE peak coincides with with a FIRST Survey radio source (Condon 1998;
Helfand \etal 2015) and an optical brightest cluster galaxy (BCG). 
It is significantly offset from the X-ray center
of the NE clump. The BCG consists of at least three
merging ellipticals and has no detected
optical emission lines (Paper I and Figure~\ref{clumpcentres}). 
The NE subcluster is itself, likely the result of a merger.
The SW component is more compact and shows a brighter 
sharp central peak. A cool core was suggested as
possible in Paper I. The BCG in the SW clump also shows no optical emission 
lines. A further comparison of the optical and X-ray appearances of the 
two components is given in Section~2.6.

\begin{figure*}
\epsscale{1.0}
\plottwo{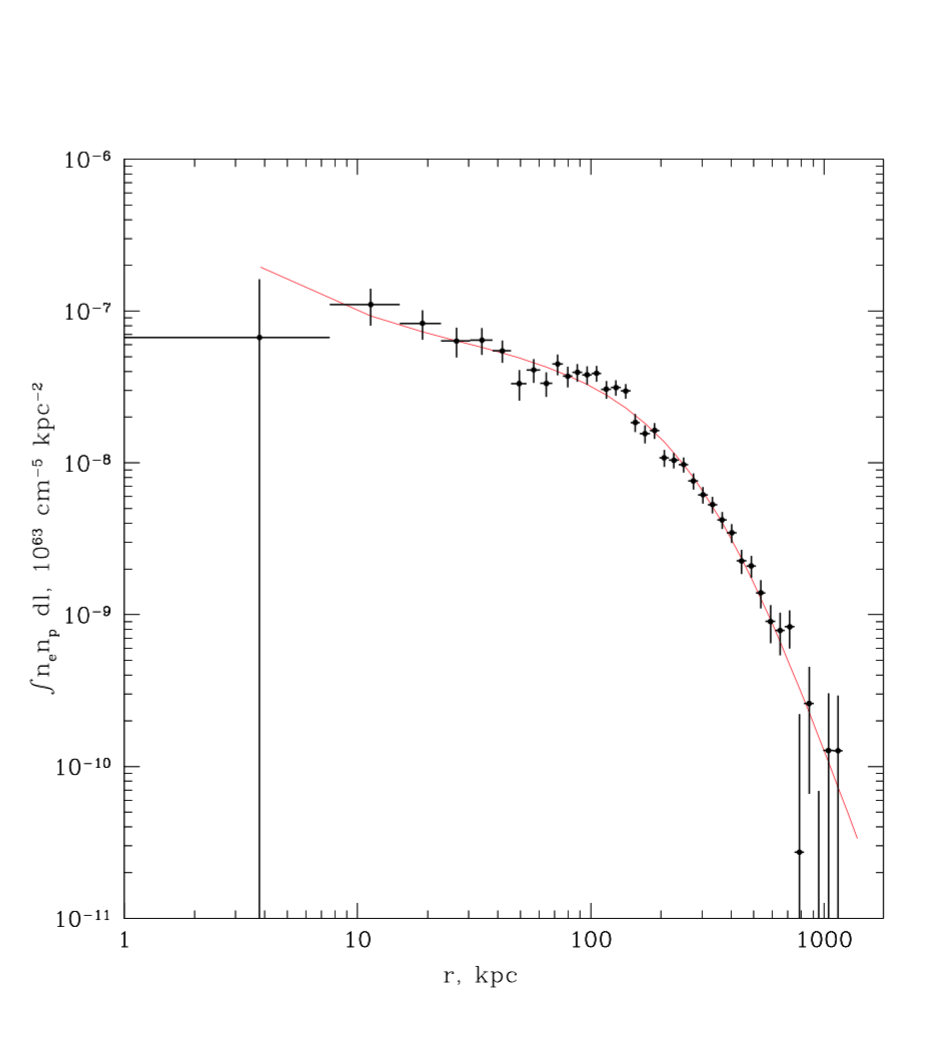}{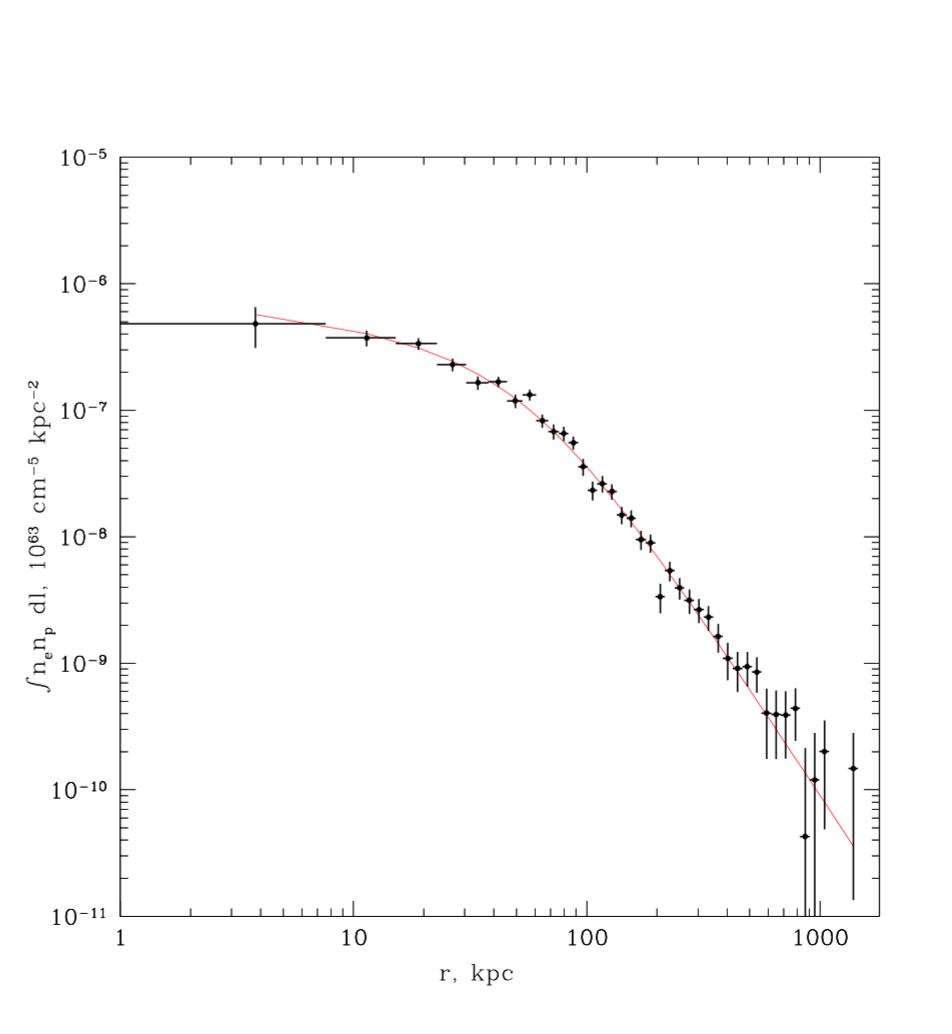}
\caption{Emission measure profiles for the NE (left) and SW (right)
 subclusters of Abell 2465 showing the \Chandra measurements and their
 error bars with the resulting model fits from Eq. (1).
}
\label{emm_profiles}
\end{figure*}

\subsection{The X-ray radial brightness and gas density profiles of Abell 2465
NE and SW} 

The projected X-ray surface brightness distributions of the 
two components of Abell 2465 were measured, from which the total X-ray 
luminosities were found to be:
$L_{X,\mathrm{bol}} = 9.0^{+0.3}_{-0.5} \times 10^{43}$\,erg\,s$^{-1}$ for the NE
component and
$L_{X,\mathrm{bol}} = 6.8^{+0.2}_{-0.3} \times 10^{43}$\,erg\,s$^{-1}$
for the SW component, comparable to the results in Paper I
from \XMM and ROSAT data. 

Assuming spherical symmetry, inverting Abel's integral, yields the emission
measure, $\epsilon_\nu$, from which results the particle densities, 
$n_p n_e = \epsilon_\nu/\Lambda(T_\mathrm{g})$ of each subcluster, where
$\Lambda(T_\mathrm{g})$ describes the emissivity of the gas at
temperature  $T_\mathrm{g}$. 

The modified $\beta$ model (Vikhlinin \etal 2006)  which fits 
the emission measure profiles of a wide range of clusters was used:
\begin{eqnarray}
 n_pn_e &=&
  n_0^2\frac{(r/\rc)^{-\alpha}}{[1+(r/\rc)^2]^{3\beta-\alpha/2}}\frac{1}{[1+(r/\rs)^\gamma]^{\epsilon/2}}
  \nonumber\\
 &&+\frac{n_{02}^2}{[1+(r^2/r_\mathrm{c2})^2]^{3\beta_{2}}}, 
\end{eqnarray}
where $n_0$ the central number density, $\beta$, and $\rc$ the core
radius have their usual meanings.  
The additional parameters, 
$\epsilon$, $\rs$, and $\gamma$ account for a slope change and 
the slope width transition.
A second small $\beta$ profile with parameters
$n_{02}, r_\mathrm{c2},$ and $\beta_{2}$ is added.
The resulting emission measure profiles of both subclusters and the fits to
(1) are shown in Figure~\ref{emm_profiles}. Table~\ref{table1} gives the 
best fit parameters for the $\beta$-model for
the NE and SW clumps. This confirms the visual impression of the differences
between the two, where the core radius, $\rc$, of the NE component is nearly
$5 \times$ larger than that of its SW neighbor.

\begin{deluxetable*}{lrrrlllllll}
\tablecaption{Chandra emission measure model best fits (Eq. [1])\label{table1}}
\tablewidth{0pt}
\tablehead{
\colhead{Sub-}&\colhead{$n_0$}&\colhead{$\rc$}&\colhead{$\rs$}&\colhead{$\alpha$}&\colhead{$\beta$}&\colhead{$\gamma$}
 & \colhead{$\epsilon$} & \colhead{$n_{02}$} & \colhead{$r_\mathrm{c2}$} & \colhead{$\beta_2$}\\
\colhead{component}&\colhead{($10^{-3}$ cm$^{-3}$)}&\colhead{(kpc)}&\colhead{(kpc)}&\colhead{}&\colhead{}&\colhead{}&\colhead{}&\colhead{(cm$^{-3}$)}&\colhead{(kpc)}&\colhead{}
}
\startdata
A2465 NE& 
1.578 & 337 & 388 &0.719 &0.849 &0.500 &0.456 &$3.55 \times 10^{-2}$ &3.37  &0.502  \\
A2465 SW& 10.763& 62  & 64  &0.693 &0.543 &0.577 &0.645 &$7.09 \times 10^{-6}$&2.03 &0.503   
\enddata
\end{deluxetable*}

\begin{figure*}
\epsscale{1.0}
\plottwo{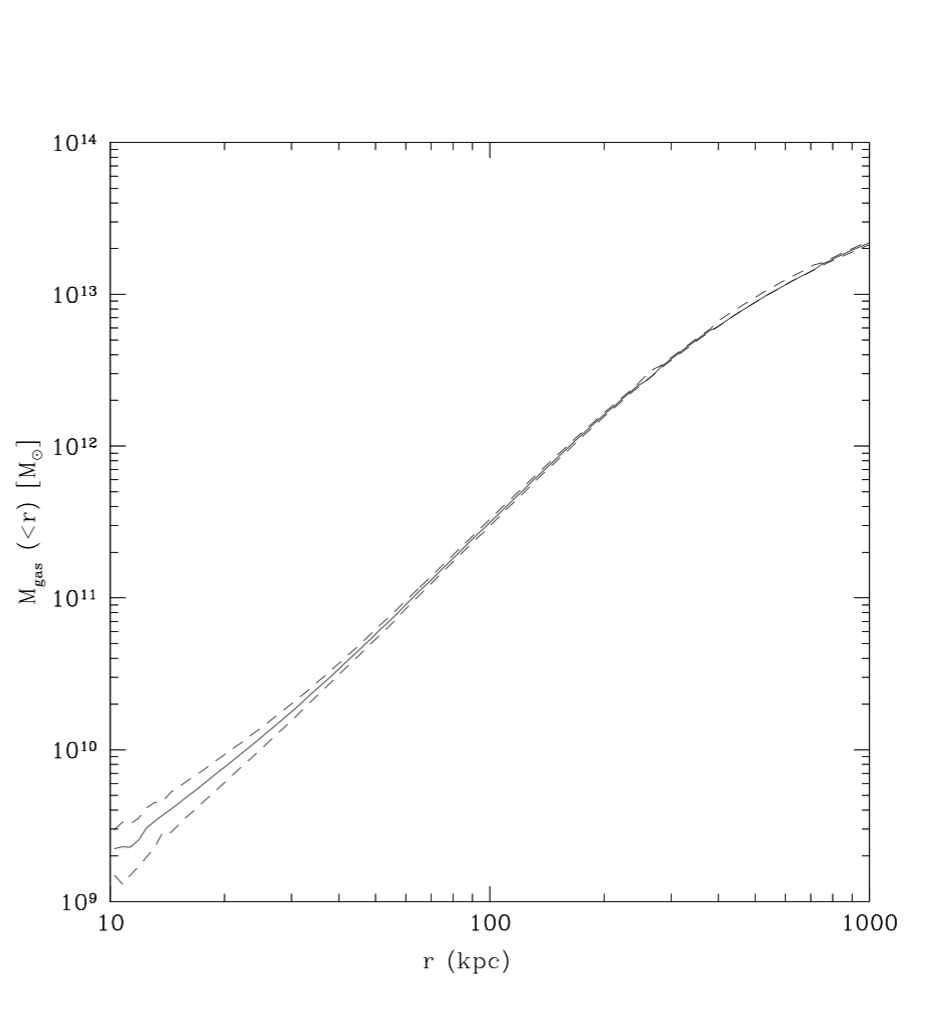}{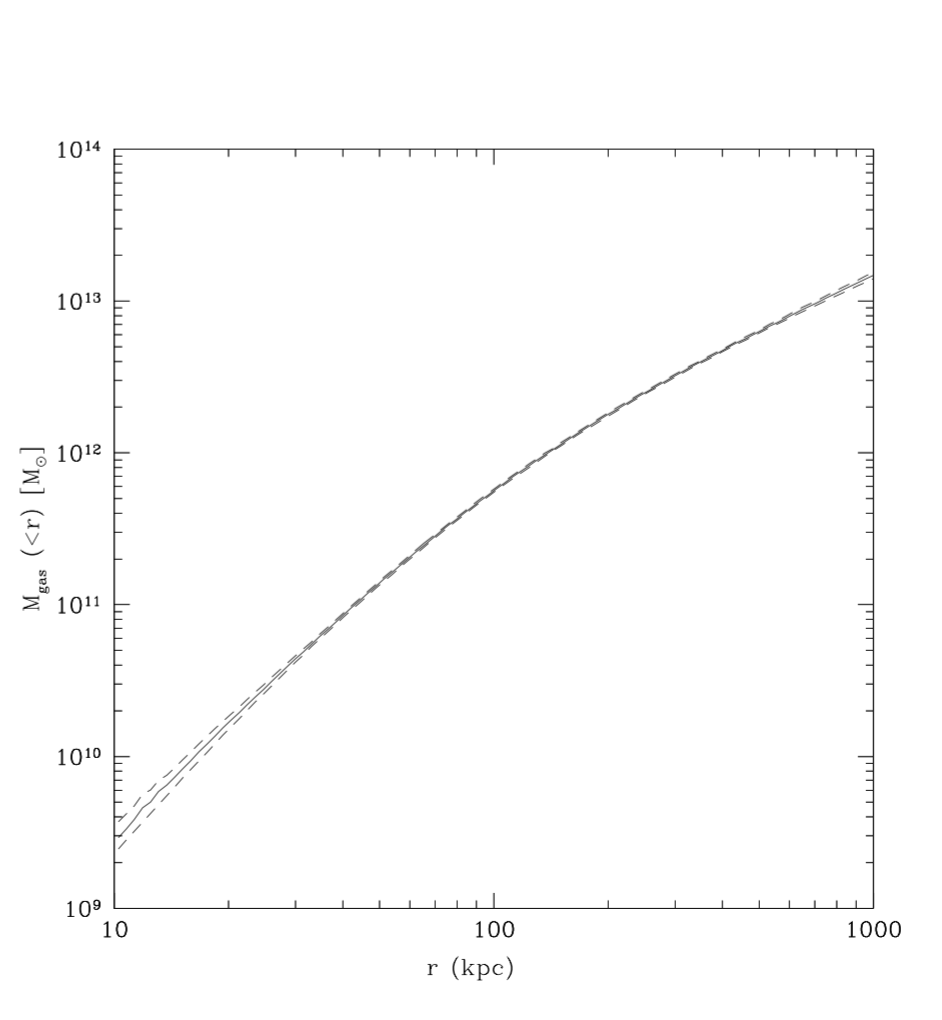}
\caption{\footnotesize Enclosed gas mass $M_\mathrm{g}(<r)$ profiles for 
 Abell 2465 NE (left) and SW (right). Dashed curves indicate $1\sigma$
 or 68\% confidence levels. 
}
\label{massprofiles}
\end{figure*}

\begin{deluxetable*}{rrrrrrrr}
\tablecaption{Fitted $\beta$-model parameters and central density\label{table2}}
\tablewidth{0pt}
\tablehead{
 \colhead{Sub-} & \colhead{$\rc$} & \colhead{$\beta$} & \colhead{$n_0$} &
 \colhead{$\rho_0$(10 kpc)} & \colhead{$M_\mathrm{hyd}$} &
 \colhead{$r_\mathrm{500c}$} & \colhead{$M_\mathrm{g}$}\\ 
 \colhead{component} & \colhead{(kpc)} & \colhead{} &
 \colhead{(10$^{-3}$ cm$^{-3}$)} & \colhead{($10^{-3}$ g cm$^{-3}$)} &
 \colhead{($10^{14} M_\odot$)} & \colhead{(kpc)} & \colhead{($10^{13} M_\odot$)}
}
\startdata
A2465 NE & $337 \pm 45$ & $0.85 \pm 0.11$ & $1.578 \pm 0003$& $9.0 \pm 3.3$& 1.85$^{+0.60}_{-0.40}$ &696$^{+75}_{-50}$  &1.90$^{+0.05}_{-0.04}$ \\
A2465 SW & $62 \pm 11$ & $0.54 \pm 0.03$ & $10.763 \pm 0.002$& $18.5 \pm 3.1$
&$2.17 \pm 0.22$  &731$\pm 25$  &$0.96 \pm 0.03$
\enddata
\end{deluxetable*}


The gas mass of each subcluster
\begin{equation} 
M_\mathrm{g} = 4\pi\int^{r_\mathrm{500c}}_{0} \rho_\mathrm{g}(r)r^2 dr
\end{equation}
within $r_\mathrm{500c}$, the spherical radius where the mean density, 
$\bar{\rho}=500\rho_\mathrm{c}$, with $\rho_\mathrm{c}$ the critical
density at the cluster's redshift, $z = 0.245$ was determined
using (1) and equation (4) in Andrade-Santos \etal
(2015) for the central electron density, $n_{e,0}$ for a plasma with fixed 
electron to hydrogen density, $n_e/n_H$, inside a ring of given inner and outer
radii,
then $\rho_\mathrm{g} = \mu_e n_e m_p$ where 
$\mu_e = 1.17$ is the electron mass molecular weight and 
$n_e/m_H = 1.2$.  

The 3D gas densities and temperature profiles are 
used to calculate the total masses, $M_\mathrm{hyd}(<r)$, inside a
clusterc radius, $r$ assuming hydrostatic equilibrium and a metal
abundance for the gas of $0.3Z_\odot$,
\begin{equation} 
M_\mathrm{hyd}(<r)=-3.67\times10^{13}M_\odot k T
 r\Big(\frac{d\ln\rho_\mathrm{g}}{d\ln r} + \frac{d\ln T}{d\ln r}\Big), 
\end{equation}
(\eg Sarazin 1986; Andrade-Santos \etal 2015), where $k$ is the Boltzmann
constant, $T$ is the gas temperature in K, and $r$ is in Mpc.

The $M_\mathrm{hyd}(<r)$ is used to obtain $r_\mathrm{500c}$ from
\begin{equation} 
M_\mathrm{hyd}(<r_\mathrm{500c}) = 500\rho_\mathrm{c}(4 \pi /3)r_\mathrm{500c}^3.
\end{equation}

\begin{deluxetable*}{lclccl}
\tablecaption{Temperature Measurements in Abell 2465\label{table3}}
\tablewidth{0pt}
\tablehead{
\colhead{Component}&
\colhead{$r$~(kpc)}&
\colhead{$kT$~(keV)}&
\colhead{Component}&
\colhead{$r$~(kpc)}&
\colhead{$kT$~(keV) }
}
\startdata
NE core& 0 - 134 & $3.38^{+0.42}_{-0.28}$& SW core  & 0 - 77 & $2.77^{+0.21}_{-0.18}$ \\
NE periphery &134 - 326 &$3.55^{+0.46}_{-0.41}$& SW periphery& 77 - 287 & $2.48^{+0.24}_{-0.23}$ 
\enddata
\end{deluxetable*}

Table 2 gives the best fitting parameters for
$M_\mathrm{hyd}, M_\mathrm{g}$, $r_\mathrm{c}$ and $\beta$
from the combined data plus the central values of the electron density
and the mass density. Within $r_\mathrm{500c}$, the two clusters have nearly equal
$M_\mathrm{hyd}$. The NE cluster has the higher gas content 
($\sim 2 \times M_\mathrm{g}/M_\mathrm{hyd}$) due
to its larger core radius and at $\rhalo$ it has the stronger $L_X$ and $\Mhalo$.

\subsection{Gas temperatures in Abell 2465}

The gas temperature of the sub-clusters could only be extracted 
near their centers.
With \Chandra, about 1800 net counts were collected from each 
sub-cluster. For the NE subcluster, a
circle (0--35)\,arcsec corresponding to (0--134)\,kpc was used; the 
corresponding area for the SW component was (0--20)\,arcsec or
(0--77)\,kpc.   
Data in the 0.6--10\,keV band were fit to an {\sc apec} single
temperature model (Smith \etal 2001).
The metallicity was assumed to be $0.3Z_\odot$. The absorption correction was
obtained from the $N_H$ value from radio surveys (Dickey \& Lockman 1990). 
The central temperatures are given in Table~\ref{table3}.
Noting the difference in their brightness profiles, the NE subcluster 
has the higher temperature indicative of a non-cool core, while the 
lower temperature of the more 
centrally concentrated SW component suggests a cool core.

\subsection{Entropy of the two sub-clusters}
The entropy provides information on a galaxy cluster's 
history (Voit \etal~2005; McDonald \etal~2013; Andrade-Santos \etal~2015). As 
the temperature profiles in Abell 2465 cannot be determined,
the entropy is limited to the central regions, within radii
0-20 arcsec.

The specific entropy is $K = kT/n_e^{2/3}$, with
$k$ the Boltzmann constant,
$T$ the temperature of the intercluster gas,  and
$n_e$ the electron density. The central entropies are found to be
$K = 78 \pm 21$\,keV\,cm$^2$ for the NE component and
$K = 40 \pm 5$\,keV\,cm$^2$ for the SW component.
For the SW sub-cluster, 
$K$ is consistent for
a $T \sim 3$\,keV relaxed cluster
where $K_0 \sim 10-30$\,keV\,cm$^2$   
(Voit \etal 2005; Cavagnolo \etal 2009; McDonald \etal 2013), 
for the NE component, $K_0$ is a factor of $\sim2$ higher.

The reduced entropy relation predicted from purely gravitational heating
is (Pratt \etal 2010),  
\begin{equation}
K/K_\mathrm{500c} = 1.42(r/r_\mathrm{500c})^{1.1}
\end{equation}
where
$K_\mathrm{500c} = 106[(M_\mathrm{500c}/10^{14}M_\odot)/f_\mathrm{b}/E(z)]^{\frac{2}{3}}$\,KeV\,cm$^{-2}$.
For both components of Abell 2465, $M_\mathrm{500c} \approx 2 \times 10^{14}M_\odot$.
With a baryon fraction $f_\mathrm{b} \approx 0.15$,
$r_\mathrm{500c} \approx 0.7$\,Mpc, 
and $E(z) = 1.131$,
$K_\mathrm{500} \approx 550$\,keV\,cm$^2$.

Galaxy mergers, AGN activity, or subcluster mergers might provide heating. 
Andrade-Santos \etal (2015) reviewed galaxy clusters with higher entropy
showing signs of merging activity with two or more core
ellipticals (\eg Cavagnola \etal 2009; Seigar \etal 2003; 
and Wang, Owen, \& Ledlow 2004).
Pratt \etal (2010) show that $K/K_\mathrm{500c}$ divides the clusters roughly into
two types which places the SW clump ($K/K_\mathrm{500c} \sim 0.07$) among cool
core clusters while the NE ($K/K_\mathrm{500c} \sim 0.14$) lies with
disturbed clusters. 
The NE clump may still be merging. In the optical, it has several BCGs
shown below in Figure~\ref{clumpcentres}. A typical 
merger time is of order a Gyr (Seigar \etal 2003).

If a shock wave raised $K$ for Abell 2465 NE, the initial $K_\mathrm{i}$
to final $K_\mathrm{f}$ entropy ratio gives the order of magnitude of
the required Mach number, $M = u/c_\mathrm{s}$, 
based on the Rankine-Hugoniot jump conditions (Andrade-Santos 2015;
Belsole \etal 2004; Zel'dovich \& Raizer 1967).  
If the NE component was initially relaxed,
$K_\mathrm{i} \sim 25$\,keV\,cm$^2$ then $K_f/K_\mathrm{i} \sim 1.6$
and $M \approx 3.28$.
For $T \sim 3$\,keV, $u \sim 2900$\,\kms; compared to
$(2G\Mhalo/\rhalo)^{\frac{1}{2}} \sim 3000$\,\kms, 
the maximum collisional velocity expected for the clusters.
This would seem possible if it were not for the lack of other
collisional signs bridging the two components.
 

\subsection{Search for inter-component gas and surface brightness jump}

\begin{figure*}
\epsscale{1.1}
\plottwo{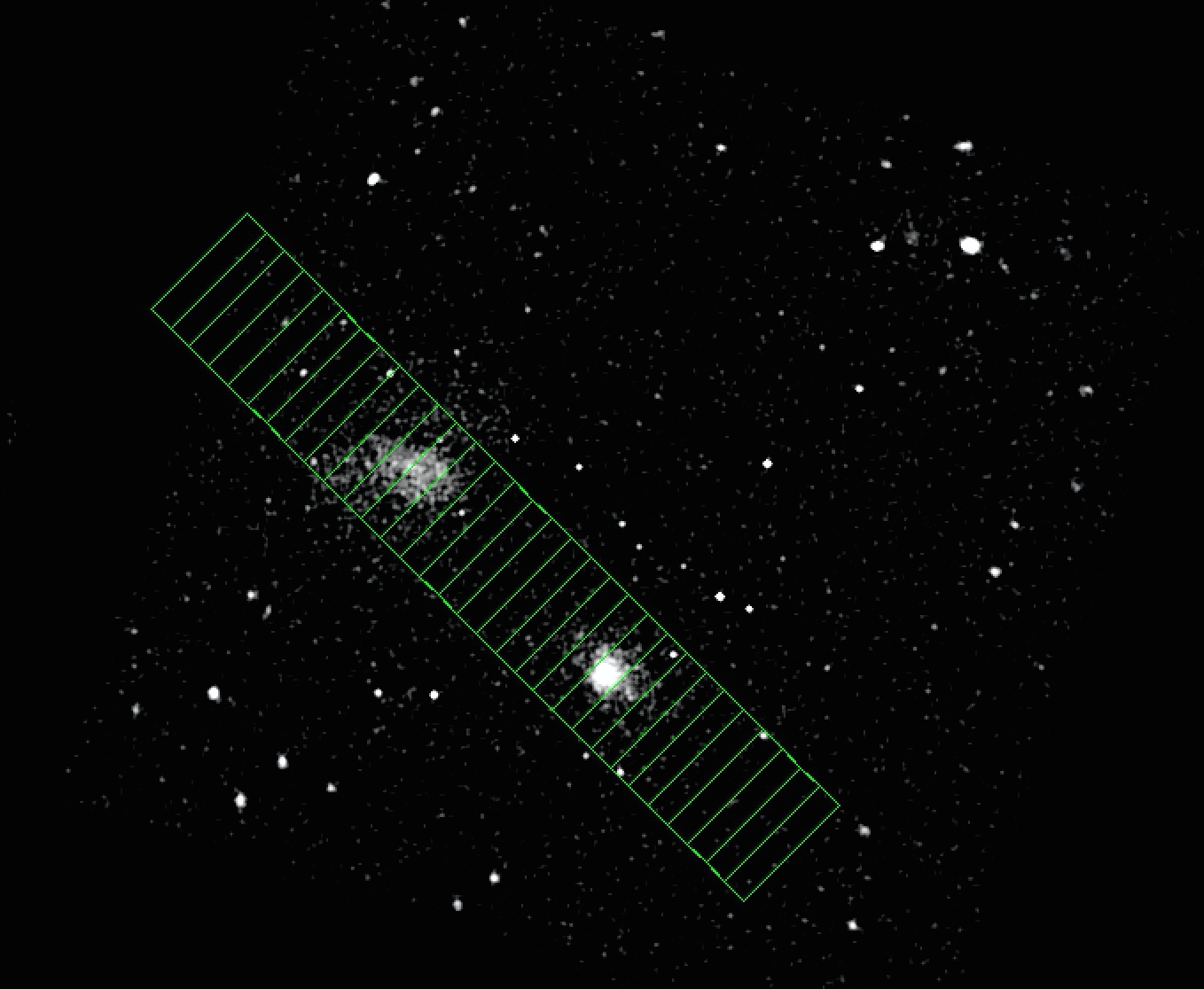}{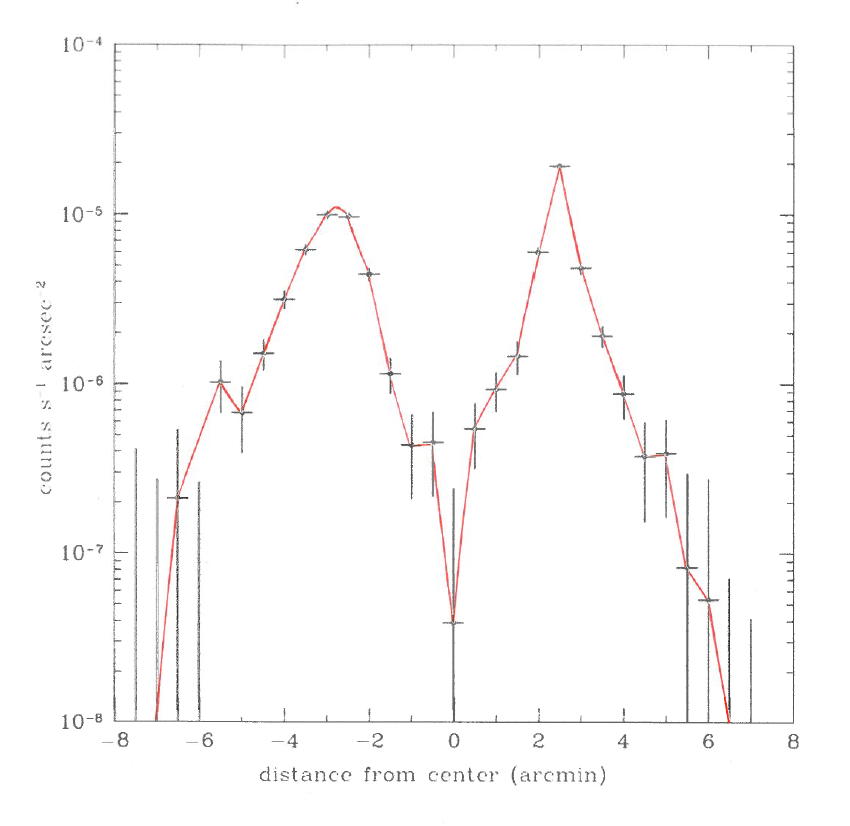}
\caption{Showing the absence of gas enhancement  and sharp intensity
 jumps between the components of Abell 2465.
 Left: locations of bins on the \Chandra (0.5--2.0)\,keV
 image; bins are $150 \times 30$\,arcmins.
 Right: surface brightness in the bins; NE sub-cluster is to the left
 and SW to the right. The peaks are 5.25\,arcmin or 1.2\,Mpc apart.
}
\label{sb_boxes}
\end{figure*}

Enhanced gas density or surface brightness jumps
between the components of Abell 2465 would be clues about
the dynamical state of the merger. We constructed a surface brightness
profile between the sub-clusters using
the 0.5--2.0\,keV \Chandra image to maximize S/N 
employing the {\sc Proffit} software package (Eckert, Molendi, \&
Paltani 2011).  
Boxes projected on the sky were
$150 \times 30$\,arcsec with a $45\degree$ position angle, and span both 
sub-clusters including regions beyond them. The surface
brightness profile in Figure~\ref{sb_boxes} reveals a hot gas 
distribution virtually identical in both directions, with no visible 
enhancement between the two sub-clusters. No statistically significant
sharp surface brightness (or density) jumps between the two clumps
were detected, and no hint of elevated brightness between them,
compared to the cluster outskirts, are detected at the
signal-to-noise of the current data. We will carry out a quantitative analysis
using numerical simulations in Section~4. 

\subsection{Optical appearance of the central BCG regions}

\begin{figure*}
\epsscale{1.0}
\plottwo{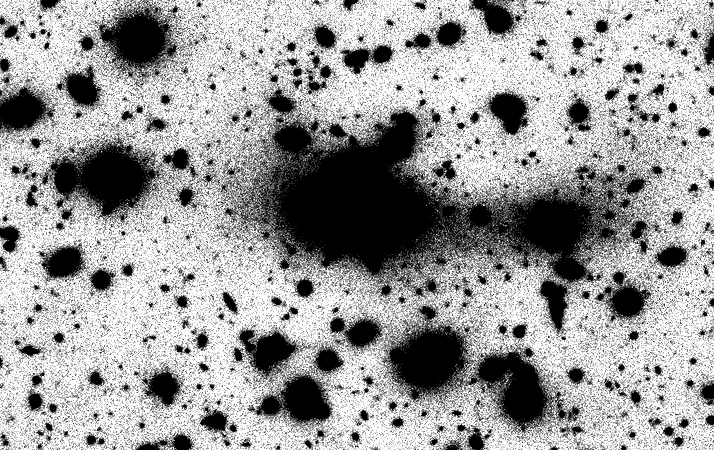}{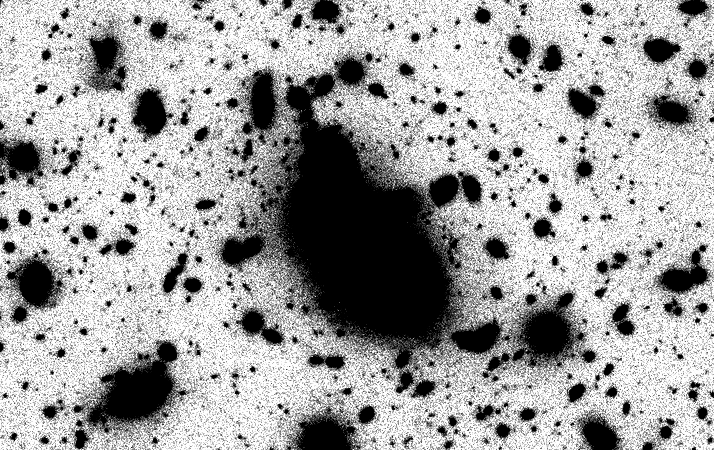}
\plottwo{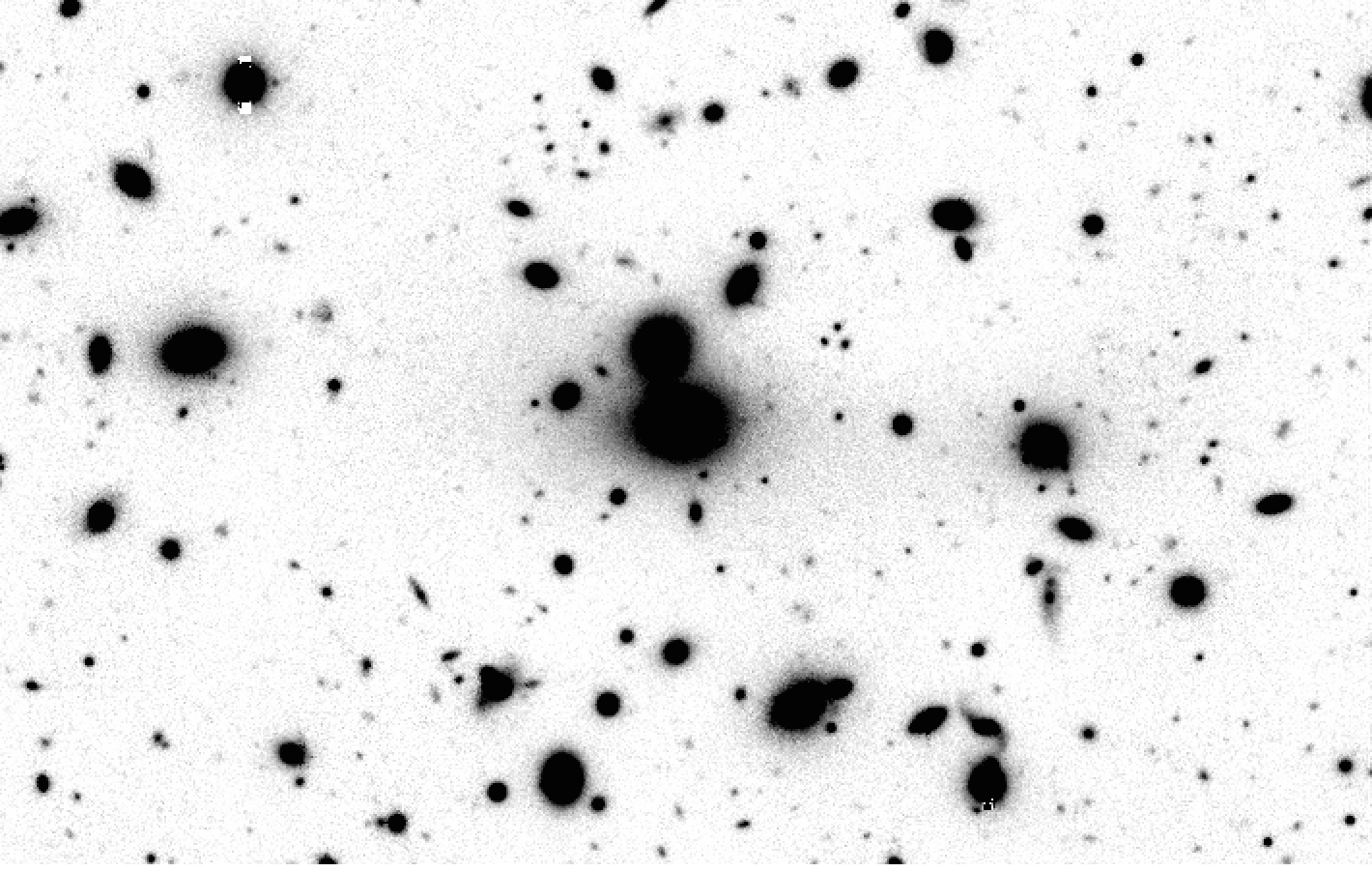}{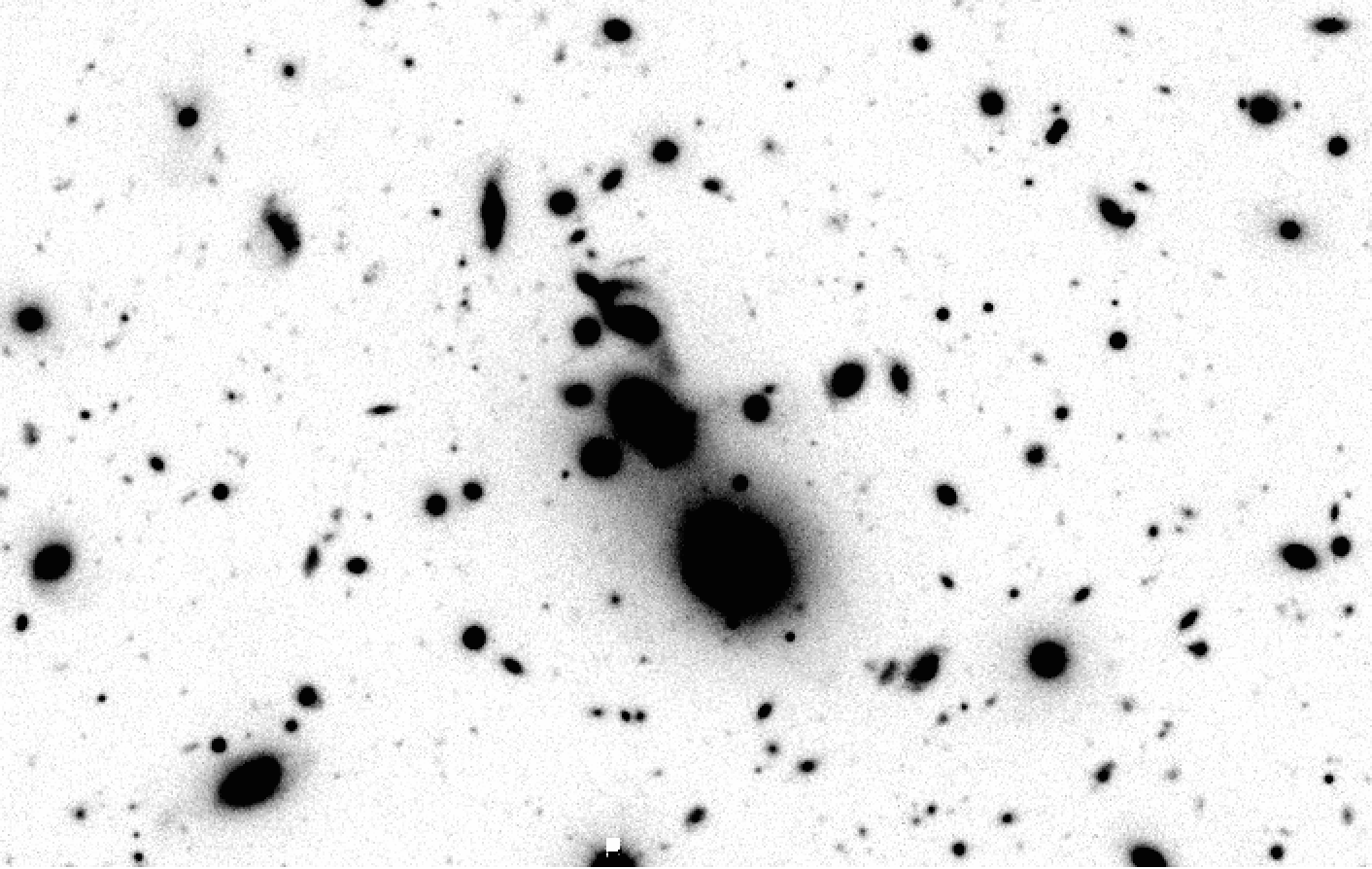}
\caption{The centers of the subcomponents of Abell 2465 from the
 combined Suprime-Cam $Vi'z'$ images described in Section 3 showing  
 the BCGs. The NE (left) component and the SW (right) component are shown.
 North is to the top and west to the right.
 The vertical edge of each image is about 1.5\,arcmin or 345\,kpc. The
 top row shows faint details near the sky limit. The bottom row is set
 to bring out the peaks of the brightest objects.
}
\label{clumpcentres}
\end{figure*}

Figure~\ref{clumpcentres} shows the centers of the NE and SW components of 
Abell 2465 in the Subaru $Vi'z'$ image described in Section 3. In 
the NE clump, 
the light of the two central E galaxies seen in Figure~\ref{isodensity} 
appears merged and
connected to a third galaxy to the W and surrounded by at least three
nearby E galaxies. The SW clump does not show such an extended distribution,
but the large central BCG is associated with several close smaller galaxies
to the north.

\section{Weak-lensing analysis}
\label{sec:wl}

Our weak-lensing results are based on Subaru/Suprime-Cam images. 
The weak-lensing methodology has been described in our previous papers
(see Umetsu \& Broadhurst 2008; Umetsu \etal 2009, 2010, 2012, 2014; 2015; 
Medezinski \etal 2013, 2016).
We therefore refer the reader to these papers and briefly outline the
methodology here.

In this work, we study the projected mass distribution in the field
of Abell 2465,
$\kappa(\btheta) = \Sigma(\btheta)/\Sigma_\mathrm{c}$,
which describes the projected mass density $\Sigma(\btheta)$ in units of
the critical surface density for lensing, 
$\Sigma_\mathrm{c} = (c^2D_\mathrm{s})/(4\pi G D_\mathrm{l}D_\mathrm{ls})=c^2/(4\pi G D_\mathrm{l}\beta)$,
where $D_\mathrm{l}, D_\mathrm{s},$ and $D_\mathrm{ls}$ are the angular
diameter distances to the lens, the source, and the lens-source,
respectively; $\beta(z,z_\mathrm{l})$ is the geometric lensing strength
as a function of source redshift $z$ and lens redshift $z_\mathrm{l}$. 

The complex gravitational shear field $\gamma(\btheta)$ is nonlocally
related to the convergence by $\partial^*\partial\kappa(\btheta)=\partial^*\partial^*\gamma(\btheta)$,
where $\partial:=\partial/\partial\theta_1 + i\partial/\partial\theta_2$ is a
complex gradient operator that transforms as a vector, 
$\partial'=\partial e^{i\phi}$, with $\phi$ the angle of rotation. 
In the subcritical regime where $(1-\kappa)^2-|\gamma|^2>0$, the reduced
gravitational shear $g(\btheta)=\gamma(\btheta)/[1-\kappa(\btheta)] (<1)$ can
be directly observed from a local ensemble of image ellipticities of
background galaxies (e.g., Bartelmann \& Schneider 2001).

In our weak-lensing analysis of the Abell 2465 field, 
we calculate the weighted average of reduced shear on a regular
Cartesian grid of cells 
($m=1,2,...,N_\mathrm{cell}$) as
\begin{equation}
\label{eq:gsm}
\langle g(\btheta_m)\rangle=\frac{\sum_i S_i(\btheta_m-\btheta_i) w_i
 g_i}{\sum_i S_i(\btheta_m-\btheta_i)w_i}, 
\end{equation}
where $S(\btheta)$ is a spatial window function, $g_i$ is the estimate
for the reduced shear of the $i$th galaxy at $\theta_i$, and $w_i$ is
the statistical weight for the $i$th galaxy,
\begin{equation}
 w_i=\frac{1}{\sigma_{g,i}^2+\alpha_g^2},
\end{equation}
with $\sigma_{g,i}^2$ the error variance of $g_i$ and $\alpha_g^2$
the constant variance taken to be $\alpha_g=0.4$, which is a typical
value of the mean rms $\sqrt{\overline{\sigma^2}_g}$ found in Subaru
observations (e.g., Umetsu \etal 2009, 2014; Okabe \etal 2010; Okabe \&
Smith 2016). The variance of the grid reduced shear is estimated by
(Umetsu \etal 2009, 2015)
\begin{equation}
\label{eq:sigma_gsm}
 \sigma^2_g(\btheta_m) = \frac{\sum_i
  S^2(\btheta_m-\btheta_i)w_i^2\sigma_{g,i}^2}
  {\left(\sum_i S(\btheta_m-\btheta_i)w_i\right)^2}.
\end{equation}

\subsection{Observational Data and Photometry}
\label{subsec:subaru}

For the weak-lensing analysis, we use archived imaging from the
Suprime-Cam ($34\arcmin\times 27\arcmin$; Miyazaki \etal 2002) at the
prime focus of the 8.3\,m  Subaru telescope, where archived images were
obtained from SMOKA  \footnote{http://smoka.nao.ac.jp}. We also included
observations from the CFHT/MegaCam.
The CFHT/MegaCam $i'$ and $r'$ images have been  describes in Paper I
and also archived. 

The imaging data are summarized in Table~\ref{obsdata}.
The reduction procedure is based on Nonino \etal (2009) and further described
in detail by Medezinski \etal (2013).
We note that the present analysis used the CFHT imaging only 
for the catalog making and magnitude zero-point calibration.

\begin{deluxetable*}{cccccl}
\centering
\tabletypesize{\scriptsize}
\tablecolumns{6}
\tablecaption{
Optical imaging data of Abell 2465
\label{obsdata}}
\tablewidth{0pt}
\tablehead{
 \colhead{Telescope}&
 \colhead{Filter}&
 \colhead{Exposure time}&
 \colhead{Seeing}&
 \colhead{$m_\mathrm{lim}$}&
 \colhead{Obs. Date}\\
 \colhead{ }  & 
 \colhead{ }  &
 \colhead {(s)} &
 \colhead {(arcsec)}&
 \colhead {(AB mag)}&
 \colhead {yy/dd/mm}
 }
\startdata
 Subaru/S-Cam& $V $& 7800 & 0.5 & 27.7 & 2006/08/26 \\     
 Subaru/S-Cam& $i'$& 3557 & 0.7 & 26.6 & 2006/08/25-26 \\ 
 Subaru/S-Cam& $z'$& 4720 & 0.8 & 26.5 & 2006/08/26 \\     
 CFHT/MegaCam& $g'$& 2060 & 0.9 & 26.1 & 2011/05/09 \\
 CFHT/MegaCam& $r'$& 1500 & 1.0 & 25.4 & 2009/17/09 \\
 CFHT/MegaCam& $i'$& 1895 & 0.5 & 25.5 & 2009/23/09 
\enddata
\end{deluxetable*}

Catalogs of objects in the images were extracted
from the available MegaCam and Suprime-Cam images. The photometric zero points
were derived using the {\sc Sextractor} program (Bertin \& Arnouts 1996) 
by matching with stars in a range of magnitude and full width half maximum
({\sc fwhm}). For the CFHT images, 
$g'$, $r'$, and $i'$
data from the Sloan Digital Sky Survey (SDSS) Data Release Nine (DR9) were used (Ahn
\etal 2012). 
For Subaru images, $z'$ zero points were estimated
from SDSS DR9, $i'$ employed the corrected CFHT $i'$ imaging above, and $V$ 
used data from Pickles \& Depagne (2010). 

For the star-galaxy separation, a plot of {\sc fwhm} versus {\sc
mag\_auto} was made to locate where the point sources lie and then
followed by a plot of
{\sc flux\_radius} versus {\sc mag\_auto}, further finalizing the 
selection. The {\sc mag\_aper} was computed for $\sim 1\arcsec$ {\sc fwhm} 
and an aperture correction was derived
from point sources to $\sim 5-10$ times {\sc fwhm} and used to recover flux 
inside 1\,{\sc fwhm} lost due to point spread function (psf) effects.

\subsection{Shape Measurements}
\label{subsec:shape}

For shape measurements, we follow the methods of Umetsu \etal (2010,
2012, 2014, 2015).
Our weak-lensing shape analysis uses the procedures of Umetsu \etal
(2014; Section 4)
employed for the CLASH survey. Briefly summarizing, the analysis
procedures include 
(see also Section 3 of Umetsu \etal 2016): 
(1) object detection using the IMCAT peak finder, {\sc hfindpeaks},
(2) careful close-pair rejection to reduce the crowding and deblending
effects,
and
(3) shear calibration developed by Umetsu \etal (2010).
We include for each galaxy a shear calibration factor of $1/0.95$ 
to account for the residual correction estimated using 
simulated Subaru/Suprime-Cam images.
To measure the shapes of the background galaxies, we use the Subaru $i$
data which have the best image quality.

\subsection{Background Selection}
\label{subsec:back}

The background selection is critical for the weak-lensing analysis
because contamination by unlensed objects will dilute the signal,
particularly at small cluster radii (Medezinski \etal 2010; Okabi \etal
2010). 

\begin{figure}
\epsscale{1.2}
\plotone{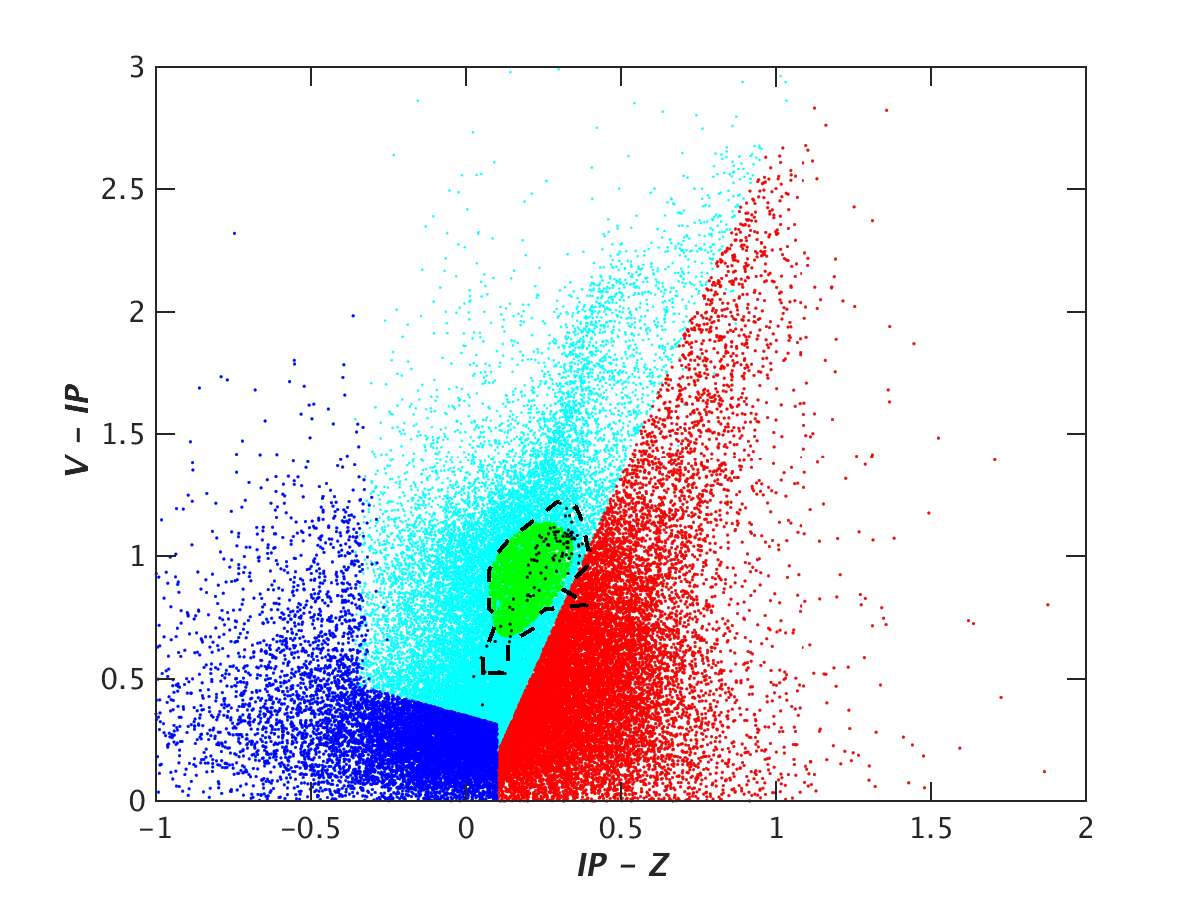}
\caption{Showing the selection of background galaxies for the WL analysis,
using Subaru $V, i'$, and $z'$ color-color selection. "Blue" galaxies
 are lower left and "red" galaxies are lower right. Rejected galaxies
 are shown as cyan. Black dots represent spectroscopically measured
 cluster members. Galaxies at small cluster-centric radius, outlined by
 dashes and colored green have been excluded from the analysis.
}
\label{selectionplot}
\end{figure}

The {Subaru $(i'-z')$--$(V-i')$} two-color diagram used for the
selection of the background 
galaxies, is shown in Figure~\ref{selectionplot}, following the background
selection method in Medezinski \etal 
(2010; 2013; 2015) and detailed in Medezinski \etal (2010). 
Selected background galaxies are shown by their respective 
blue and red colors. The region occupied by the spectroscopic sample of
cluster members is outlined by the black dashed curve and the cluster
member region is green.  

The background sample contains 9198 (blue + red) source galaxies,
corresponding to the mean surface number density of  
$\overline{n}_g\simeq 15$\, galaxies\,arcmin$^{-2}$.
We estimate the mean depths $\langle \beta\rangle$ of the blue and red
background samples, which are needed when
converting the measured lensing signal into physical mass
units. 
To this end, we rely on accurate photometric redshifts derived for
COSMOS (Capak \etal 2007) by Ilbert \etal (2009) using 30 bands in the
ultraviolet to mid-infrared. We apply the same color-color/magnitude limits as
for A2465 on the COSMOS catalog. However, since COSMOS photometric
redshifts are reliable only to a magnitude of $\lesssim25$, whereas
Subaru is deep to $\lesssim 25.4$, we limit the redshift estimation to
$z<4$, and to magnitude $z'<25$, and extrapolate the relation between
depth and magnitude $\langle\beta\rangle(z')$ further from $25<z'<26$. Using COSMOS
we calculate the depth of each sample, red and blue, separately. Finally
we derive the mean value taking into account the relative fraction of
red and blue galaxies of the total Subaru red+blue sample. 
For the composite blue+red sample, we find 
$\langle \beta \rangle = 0.7130 \pm 0.036$, corresponding to an effective 
source redshift of $z_\mathrm{eff}\simeq 1.06$.


For weak-lensing mapmaking, we draw a looser background sample,
comprised of all the galaxies outside the region defined by the
spectroscopic members (dashed black curves). This sample has a mean
source density of $\overline{n}_g\simeq 28$\,galaxies\,arcmin$^{-2}$,
which is about $90\%$ higher than that of the stringent background
selection.

In Figure~\ref{fig:map}, we show a {Subaru $Vi'z'$} composite image of the
cluster field, produced using the publicly available {\sc trilogy}
software (Coe \etal 2012). The image is $14\arcmin\times 14\arcmin$ in
size and overlaid by Gaussian-smoothed
($2\arcmin$\,{\sc fwhm}) weak-lensing mass contours for visualization purposes.
Here we used the Gaussian smoothing kernel as a filer function
$S(\btheta)$ in Equation (\ref{eq:gsm}). 

The SW mass distribution appears pulled out along the
line joining the centers of the two components. Such elongation would be 
consistent with tidal interaction between the two sublusters. In paper I, the
corresponding light distributions were examined. While the SW components has
a sharp peak, it also seems to have an extended underlying distribution of
faint galaxies and its luminosity function was found to have more galaxies
with $M_I \ga 20$ mag.

We note that, we use this sample only for visualization purposes because
it suffers from some degree of dilution. 
We will use the stringent (blue + red) background sample when fitting the
multi-halo model to the two-dimensional shear data to find masses of the
substructures (Section \ref{subsec:wlmass}).

\subsection{Weak-lensing Multi-halo Mass Modeling}
\label{subsec:wlmass}


We perform a two-dimensional shear fitting (Okabe \etal 2011; Watanabe
\etal 2011; Umetsu \etal 2012; Medezinski \etal 2013, 2016),
by simultaneously modeling the two components of Abell 2465 as a
composite of two spherical halos.

To do this, we construct pixelized maps of the two-dimensional
reduced-shear $g(\btheta)$ (Equation \ref{eq:gsm})
and its error variance $\sigma_g(\btheta)$ (Equation (\ref{eq:sigma_gsm}))
on a Cartesian grid of $20\times 24$ independent cells
($N_\mathrm{cell}=480$) with $0.75\arcmin$ spacing. Here we have adopted
bin averaging, corresponding to $S=1$ in Equation (\ref{eq:gsm}). Our
multi-halo modeling is restricted to a central region with $15\arcmin\times
18\arcmin$ that contain the NE and SW components.
To avoid systematic errors, we have excluded from our analysis innermost
cells lying at $|\btheta|<1\arcmin$ from each of the halos (Oguri \etal
2010; Umetsu \etal 2012), where the surface-mass density can be close to
or greater than the critical value,
to minimize contamination by unlensed cluster member galaxies (Section
\ref{subsec:back}) as well as to avoid the inclusion of strongly lensed
background galaxies. 

We adopt the Navarro--Frenk--White (Navarro, Frenk, \& White 1997; NFW,
hereafter) model to describe the mass distribution of each cluster
component. The NFW density profile provides a good description of the
observed mass distribution in the intracluster regime, at least in an
ensemble-average sense (e.g., Umetsu \etal 2011a, 2011b, 2014, 2016;
Okabe \etal 2013; Niikura \etal 2015; Okabe \& Smith 2016). We specify
the NFW model using the halo mass $\Mhalo$,
concentration $\chalo=\rhalo/r_{-2}$ with $r_{-2}$ the characteristic
radius at which the logarithmic density slope is $-2$, and centroid
position on the sky. 

We adopt an uninformative log-uniform prior in the halo-mass interval,
$0.1\le \Mhalo/(10^{14}M_\odot\,h^{-1})\le 100$.
We set the concentration parameter for each halo using the theoretical
concentration--mass relation of Dutton \& Maccio (2014), which is
calibrated using a {\em Planck} cosmology and is in good agreement with
recent cluster lensing observations (Umetsu \etal 2016; Okabe \& Smith
2016). 
For the halo centroid, we assume a Gaussian prior centered on each of
the BCG position with standard deviation
$\sigma=0.5\arcmin$ ($\textsc{fwhm}=1.18\arcmin$).
Accordingly, each of the NFW halos is specified by three
model parameters, $(\Mhalo,\Delta\mathrm{RA},\Delta\mathrm{Dec})$, 
where the centroid ($\Delta\mathrm{RA},\Delta\mathrm{Dec}$) is defined
relative to the BCG position. 

We use the Markov Chain Monte Carlo algorithm with Metropolis--Hastings
sampling to constrain the multi-halo lens model from a simultaneous 
six-component fitting to the reduced-shear field $g(\btheta)$. We employ 
the shear log-likelihood function of Umetsu \etal (2012; Appendix A.2)
and Umetsu \etal (2016).   

\begin{figure*}[!htb] 
 \begin{center} 
\includegraphics[width=0.8\textwidth,angle=0,clip]{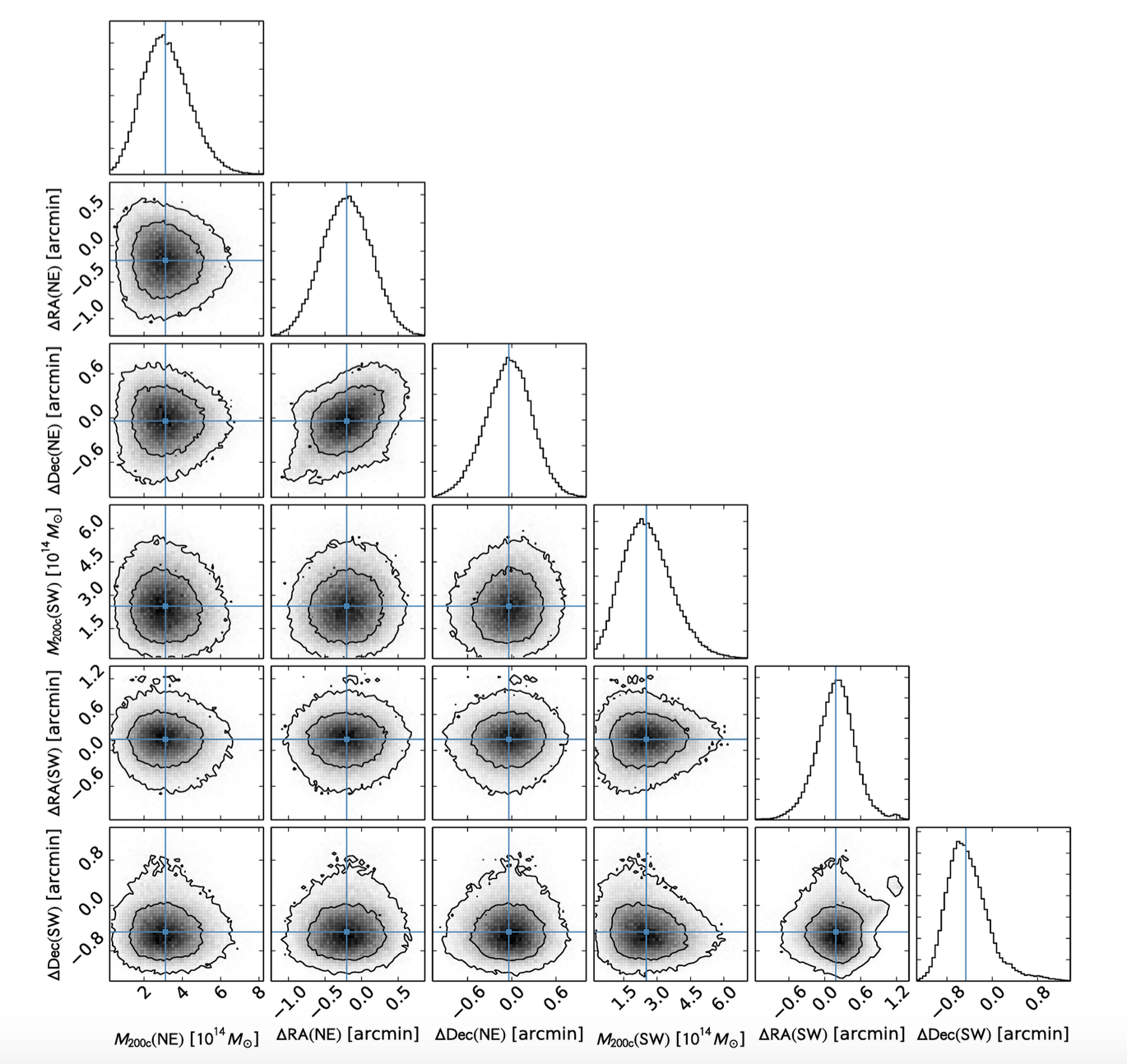} 
 \end{center}
 \caption{\label{fig:multihalos}
The posterior distributions of the NFW model parameters for the NE
and SW components of Abell 2465,
obtained from multi-halo fitting to the two-dimensional Subaru reduced-shear 
field.
For each parameter, the blue solid line shows the biweight central
 location of the marginalized 
one-dimensional posterior distribution. 
For each component, 
the halo centroid ($\Delta\mathrm{RA}$, $\Delta\mathrm{Dec}$) is measured
 relative to its brightest galaxy position.
}
\end{figure*}

The marginalized posterior distributions for the multi-halo model are
shown in Figure \ref{fig:multihalos}.  
The resulting constraints on the halo masses are summarized as follows:
\begin{itemize}
\item $\Mhalo(\mathrm{NE}) = (3.1 \pm 1.2)\times10^{14}M_\odot$  
      ($\chalo=4.40\pm 0.17$),
\item $\Mhalo(\mathrm{SW}) = (2.5 \pm 1.1)\times10^{14}M_\odot$
      ($\chalo=4.49\pm 0.20$),
\end{itemize}
where we employ the robust biweight estimators of Beers \etal
(1990) for the central location (mean) and scale (standard
deviation) of the marginalized posterior distributions (e.g.,
Sereno \& Umetsu 2011; Umetsu \etal 2014).
The resulting weak-lensing mass estimates are noisy with about 40\%
uncertainty.
The mass ratio between the SW and NE halos is constrained as
$\Mhalo(\mathrm{SW})/\Mhalo(\mathrm{NE}) = 0.82 \pm 0.51$.
At the virial overdensity, we find
\begin{itemize}
\item $\Mvir(\mathrm{NE}) = (3.7 \pm 1.5)\times10^{14}M_\odot$  
      ($\cvir=5.46\pm 0.21$),
\item $\Mvir(\mathrm{SW}) = (2.9 \pm 1.3)\times10^{14}M_\odot$
      ($\cvir=5.57\pm 0.24$),
\end{itemize}

These can be compared with the virial masses found in Paper I which 
are
$(4.1 \pm 0.8)\times 10^{14} M_\odot$
and 
$(3.8 \pm 0.8)\times 10^{14} M_\odot$ respectively.

\begin{figure*}[!htb] 
 \begin{center} 
\includegraphics[width=0.8\textwidth,angle=0,clip]{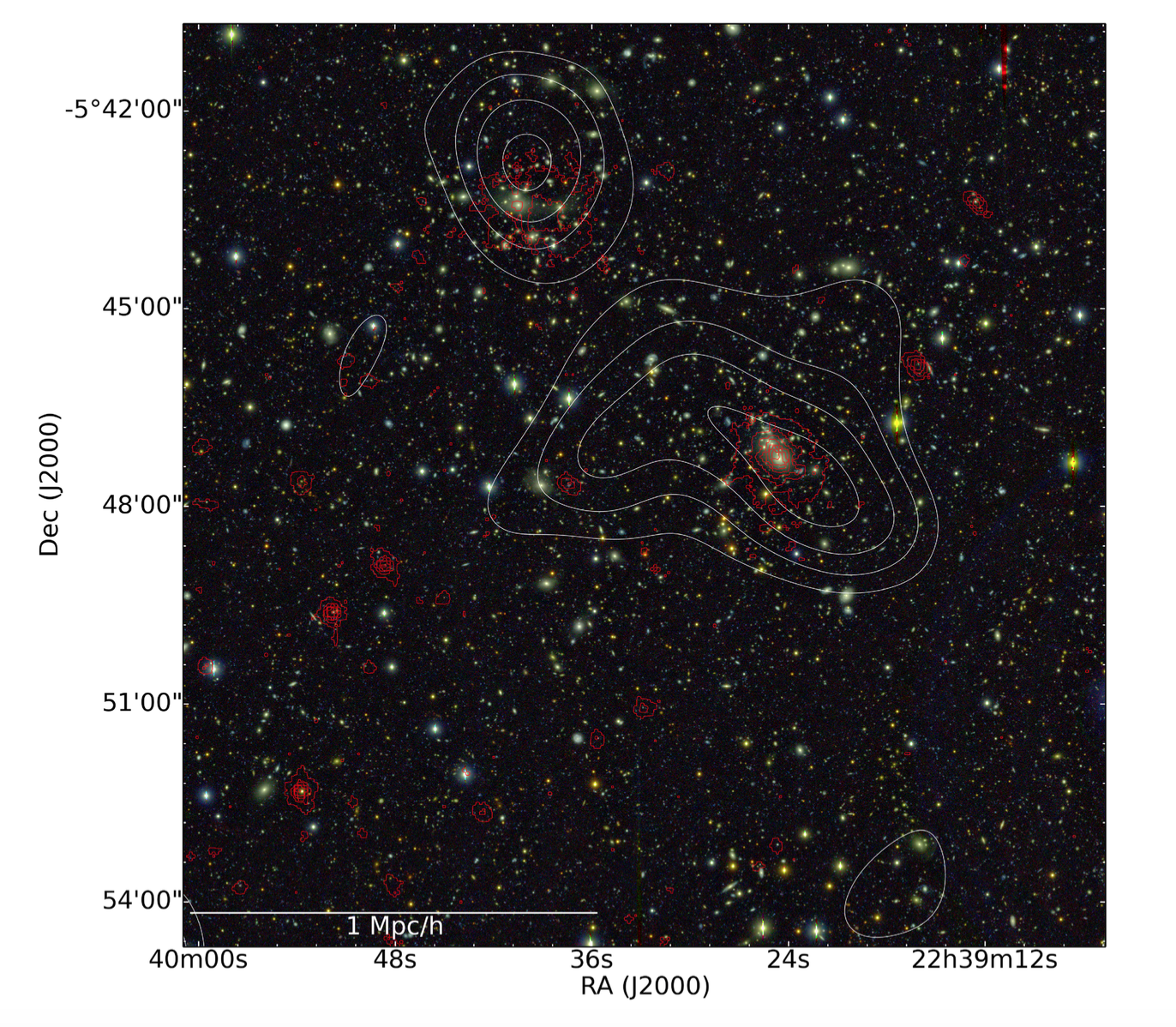}
 \end{center}
 \caption{\label{fig:map}
Subaru $Vi'z'$ composite color image in the field of Abell 2465,
overlaid with weak-lensing mass contours.
The mass map is smoothed with a $2\arcmin$\,FWHM. 
The lowest contour is at the $2\sigma$ reconstruction error level
 ($\kappa=0.034$),
and the contour interval is $\Delta\kappa=0.017$.
The image is $14\arcmin\times 14\arcmin$ in size. 
The horizontal bar represents 1\,Mpc\,$h^{-1}$ at the cluster redshift
 of $z_\mathrm{l}=0.245$. 
The red contours show the smoothed \Chandra X-ray brightness data.
North is up and east to the left.
 }
\end{figure*}


Along the line joining the NE and SW sub-clusters,
the offsets of the peaks of the X-ray distributions in 
Figure~\ref{fig:map} are about 0.9 and 0.7 arcmin (or
$\sim 0.2$ Mpc) closer together than the weak-lensing peaks. Relative to
the BCGs, the weak-lensing offsets are smaller, being about 0.5 arcmin
and the distances of the X-ray are about half of this value.   
Although this is the right order of magnitude for
the separation of dark matter and baryonic
matter in a cluster merger, given the 1.7 arcmin smoothing, this offset is 
probably insignificant.


\section{The dynamical state of Abell 2465}

The radial infall model (Beers, Huchra \& Geller 1982) 
is discussed in Paper I. 
Two mass points of total mass, $M$, are bound if 
$V_r^2R_\mathrm{p} \leq 2GM\sin^2\phi\cos\phi$. 
With $V_r = 205$\,km\,s$^{-1}$ (Paper I), Abell 2465 satisfies
this condition. Three possible solutions
depend on inclination, $\phi$, maximum orbital separation,
$R_\mathrm{m}$, the system's age, $t_0$, assumed to be the age of
the universe at redshift, $z$, and the development parameter, $\eta$.
The observed projected distance between the masses centers is 
$R_\mathrm{p} = R\cos\phi$ and the observed velocity difference is
$V_r = V\sin\phi$. 

Using $M = 8 \pm 1 \times 10^{14}M_\odot$,
$R_\mathrm{p} = 1.265$\,Mpc, and 
$t_0 = 10.895$\,Gyr in Paper I, the three solutions are:
\begin{enumerate}
\item $\eta =5.17$\,rad, $\phi = 5.95\degr$, $R=1.31$\,Mpc,
      $R_\mathrm{m}=4.53$\,Mpc, $V=-1978$\,km\,s$^{-1}$ 
\item $\eta =3.53$\,rad, $\phi = 77.5\degr$, $R=6.01$\,Mpc,
      $R_m=6.07$\,Mpc, $V=-210$\,km\,s$^{-1}$ 
\item $\eta =2.68$\,rad, $\phi = 81.3\degr$, $R=8.59$\,Mpc,
      $R_m=8.81$\,Mpc, $V=+208$\,km\,s$^{-1}$ 
\end{enumerate}
Solution (3) has the subclusters moving apart and approaching maximum
separation was favored in Paper I, but given the discussion here, 
that NE and SW have not yet collided, the other two now
seem more likely. According to solutions (1) and (2), a core passage
would occur in 0.4 or 6.6 Gyr respectively.
Possibly the (1) solution is preferable in light of the enhanced star
formation induced by the higher impact velocity.

For more detailed modelling, we 
follow Molnar \etal (2013) who analyzed the Abell 1750 double cluster. This
employs the FLASH program which is a parallel Eulerian code originating
from the Center for Astrophysical Thermonuclear Flashes at the University of
Chicago (Fryxell \etal 2000; Ricker 2008). Further details of the 
binary merger models are given in Molnar, Hearn, \& Stadel (2012). Briefly,
the dark matter and galaxies are modelled by truncated NFW profiles
and $\beta$-models are used for the gas. The velocities are taken to be
isotropic and follow the relations in Lokas \& Mamon (2001). 

\begin{figure}
\centering
\includegraphics[width=0.48\textwidth]{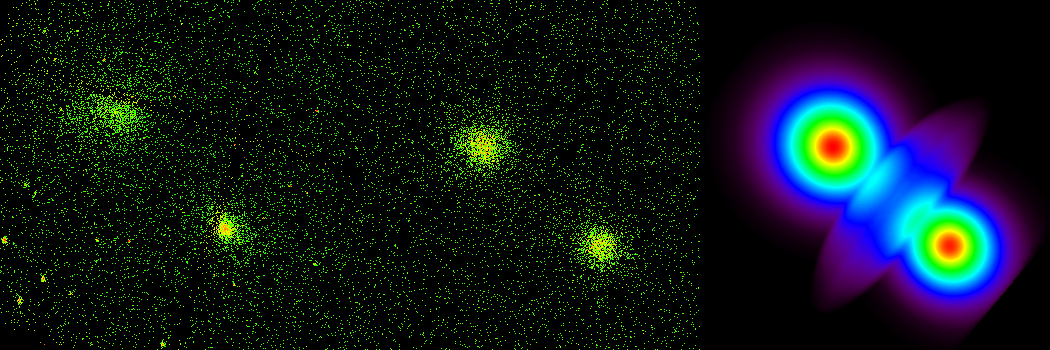}
\parbox{ \linewidth}{\hspace{0.05 cm}V$_\mathrm{infall}$ = 1000~\kms} 
\\
 \label{}
\includegraphics[width=0.48\textwidth]{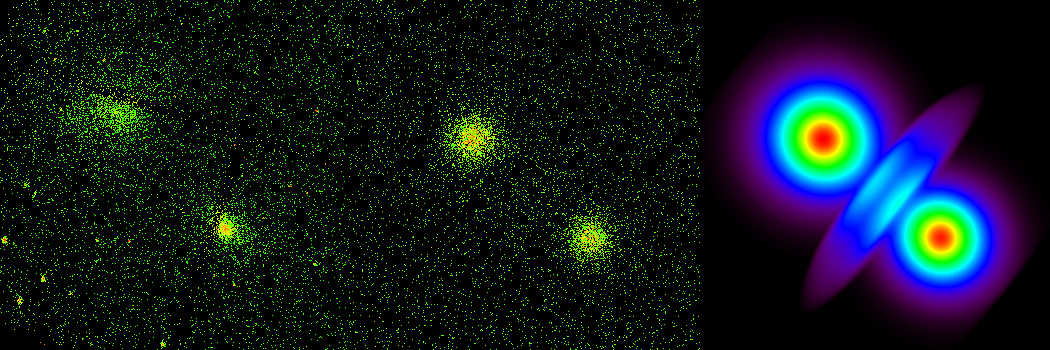}
\parbox{ \linewidth}{\hspace{0.05 cm}V$_\mathrm{infall}$ = 2000~\kms} 
\\
 \label{}
\includegraphics[width=0.48\textwidth]{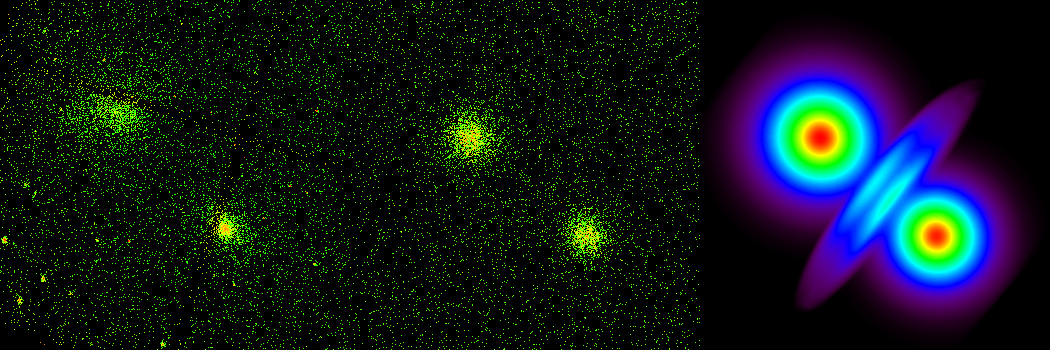}
\parbox{ \linewidth}{\hspace{0.05 cm}V$_\mathrm{infall}$ = 3000~\kms} 
\caption{Examples of simulated X-ray observations based on 
our FLASH simulations with different infall velocities as marked
(V$_\mathrm{infall}$; top to bottom) 
for Abell 2465 compared with \Chandra observations.
from right to left: 
image from \Chandra observations; 
mock X-ray observation fitted to the \Chandra flux;
model X-ray surface brightness.
}
\label{F:SIMULV123}

\end{figure}

In simulating the Abell 2465 system, we assumed 
masses,
$M_\mathrm{vir,1} = 4 \times 10^{14} M_\odot$ and
$M_\mathrm{vir,2} = 3 \times 10^{14} M_\odot$ 
and concentration parameters,
$c_\mathrm{vir,1} = 5$ and
$c_\mathrm{vir,2} = 6$,
which are consistent with the weak-lensing measurements presented in
Section~\ref{subsec:wlmass}.
We ran FLASH simulations with a range of collisional velocities,
$V_\mathrm{infall}$
and impact parameters, $P$. We generated mock X-ray observations based
on our simulations choosing the phase of the collision which match the
observations,
and added noise similar to that of the \Chandra observations of A2465.
The results presented here are relevant to our study of the
dynamical state of A2465.
Models with 
$V_\mathrm{infall}=1000, 2000$, and $3000$\,km\,s$^{-1}$ and $P=150$\,kpc
are shown in Figure~\ref{F:SIMULV123} (top to bottom).
In this figure we show cluster X-ray surface brightness distribution (no
noise) based on our FLASH simulations, images of mock \Chandra
observations (noise added),
and, for comparison, the X-ray image from our \Chandra observation (left 
to right).
It can be seen from the X-ray surface brightness distribution (right
panels), that
the morphology of the emission changes as a function of infall velocity.
The two X-ray peaks associated with the shock/compression heated
intracluster gas
of the two components are closer to the centers of the colliding
clusters for lower infall velocities.
Unfortunately the depth of the X-ray observations is insufficiently deep 
enough to see the detailed morphology of the interaction region.

Solutions (2) and (3) imply that the two clusters appear close to each
other only due to a projection effect, they are actually 6\,Mpc and
8.59\,Mpc apart, and the collision is close to the LOS
($\phi = 77.5\degr$ and $\phi = 81.3\degr$).
In this case no enhanced X-ray emission should be observed between the
two cluster centers, just a simple superposition of their equilibrium
gas emission.

Solution (1) of our simplified dynamical analysis
suggests that the collision is close to the plane of the sky,
$\phi = 5.95\degr$, and the intracluster gas of the two components
are already interacting (in collision), since $R=1.31$\,Mpc is less than
the sum of the two virial radii. In this case, we should see enhanced
X-ray emission from the shock/compression heated intracluster gas
between the two cluster centers.
In solution (1), the 3D relative velocity between the sub-clusters is
$V=-1978$\,\kms. 
Our simulations with infall velocities of
$V_\mathrm{infall} = 1000$\,\kms and 2000\,\kms
bracket this value (note that the infalling cluster speeds up as it
falls in, so its infall velocity should be slightly less than 
$V = -1978$\,\kms).
Note that at this large distance, $R\geq 1.3$\,Mpc, the relative velocity
of the infalling cluster is insensitive to the expected impact parameter
($P\simlt 300$\,kpc), since it is moving in the shallow outer part of
the cluster's gravitational potential.

\begin{figure}
\centering
\includegraphics[width=0.475\textwidth]{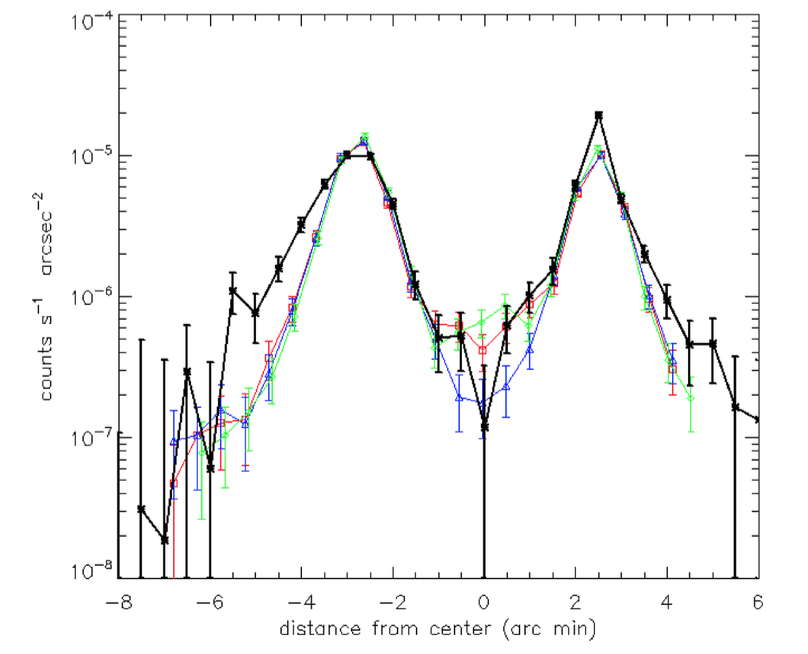}
\caption{
X-ray surface brightness profiles along the line connecting the 
two X-ray peaks from \Chandra observations
(stars with error bars connected with thick black solid lines),
from mock observations with V$_\mathrm{infall}$ = 1000 and 2000 \kms
(red squares and green diamonds with error bars connected with solid lines 
of the same color; presented in Figure~\ref{F:SIMULV123}),
and from mock observations assuming that the clusters are not interacting,
they look close to each other only in projection (blue triangles with error bars).
}
\label{F:SIMULPROF}

\end{figure}

Consequently we extract data from our mock images based on our FLASH
simulations assuming $V_\mathrm{infall} = 1000$\,\kms and 2000\,\kms
from the same regions in the sky used for our \Chandra analysis 
(shown in Figure~\ref{sb_boxes}, and derive the X-ray
profiles along the line between the two X-ray peaks.
In Figure~\ref{F:SIMULPROF} 
we compare the surface brightness profile between the
two cluster centers extracted from our \Chandra observations 
(stars with error bars connected with black thick solid line),
and the profiles extracted from our mock observations based on
simulations with $V_\mathrm{infall} = 1000$\,\kms and 2000\,\kms
(red squares and green diamonds with error bars connected with solid lines 
of the same color; same data were used in Figure~\ref{F:SIMULV123}).
Triangles with error bars connected with a blue solid line represent an
X-ray profile assuming no interaction between the two clusters 
(corresponding to solutions (2) and (3)).
As it can be seen from this figure, the simulations predict an enhancement
between the two X-ray peaks if there is a collision in progress.
The enhancement in the X-ray emission seems to be significant 
between the two cluster centers even with the low exposure time
assumed for the image simulations 
(the red squares and green diamonds in the middle region are 
about $1\sigma$ higher than the blue triangles, which represent 
on-interacting clusters).
However, we can see no significant difference between simulations assuming 
$V_\mathrm{infall} = 1000$\,\kms and 2000\,\kms with the given exposure times.
Thus our simulations suggest that A2465 is in a process of collision, the
intracluster gas of the two components are already interacting, but we cannot
constrain the infall velocity of the system using only our FLASH simulations.

\section{Discussion}


From the increasing number of examples of 
double and multiple merging galaxy clusters, 
discovered and analyzed in recent years, these objects can 
provide information on galaxy formation and evolution,
give clues on the interactions of the clusters' baryonic and dark
matter interactions, as well as the behavior of gravitation on Mpc
scales. The interactions of different types of dark matter through
pressure effects (Ota \& Yoshida 2016), dynamical friction and self
interacting dark matter (Irshad \etal 2014; Kahlhoefer \etal 2014) 
or modified gravity (\eg Del Popolo 2013; Matsakos \& Diaferio 2016)
could in principle produce observable effects.
To date, this has proven difficult due to the complexity of the
interacting systems which range from pre-mergers, given in the
Introduction (Okabe \& Umetsu 2008; Andrade-Santos \etal 2015) as well as
the A3407+A3408 pair (Nascimento \etal 2016) to post-core passage objects.
However, in many cases the structure geometry is complex. The
$\sim 2 \times 10^{15} M_\odot$
'El Gordo' cluster (Jee \etal 2016) only shows one X-ray peak, 
CIZA J2242.8+5301 is highly elongated (Jee \etal 2015), and 
CIZA J0107.7+5408 (Randall \etal 2016) is a complex dissociative merger.   

Using the basic properties of the two sub-components in Abell 2465
derived above and in Papers I and II, scaling relations for galaxy
clusters, the merger indicate no observable effects other than the
enhanced star formation found in Paper II. Both clumps follow
the cluster $\Mhalo$--$L_X$ relations (\eg Reiprich \& B\"ohringer 2002;
Rykoff \etal 2008) and the $T_X$--$\Mhalo$ relations (\eg Popesso \etal
2005). 
However, the SW subcluster appears to have a lower than
normal gas fraction, $f_\mathrm{gas} = M_\mathrm{gas}/M_\mathrm{hyd}$. From 
Table~\ref{table2}, $f_\mathrm{gas} = 0.10$ for the NE clump, and it is
0.04 for the SW clump. 
In the $f_\mathrm{gas}$--$T_X$ relations in Sanderson \etal (2003),
Vikhlinin \etal (2006), Sun \etal (2009), Sun (2012), and Lovisari \etal
(2015), the corresponding gas fractions are about 0.10.

The reduced entropy, 
$K/K_\mathrm{500c} \approx 0.07,$ for the SW component is in the range
for cool-core or gravitationally
collapsed clusters (\eg Pratt \etal 2010) and the higher value 
($K/K_\mathrm{500c} \approx 0.14$) for the NE could
be attributed to a morphologically disturbed cluster due to
a separate earlier merging event rather than interaction with 
the SW. Due to our limited data, this should only be considered
indicative rather than conclusive.

The low exposure time of our X-ray observations of A2465 does not allow
us to carry out a quantitative dynamical analysis based on
$N$-body/hydrodynamical simulations. The infall velocity, impact
parameter, \etc of the system can not be
derived, but combining a simplified dynamical analysis with FLASH
simulation allows us to find an approximate dynamical state of the
system, specifically, to distinguish between a non-interacting system
and one in an early state of merging.
Our FLASH simulations suggest enhanced X-ray emission between
the two cluster centers (relative to assuming a simple superposition of
the two equilibrium cluster emission), which can be attributed to
shock/compression heated intracluster gas due to merging.
We have found only one solution, applying our simplified dynamical analysis,
which results in a stage of merging, close to solution (1) 
of the radial infall model with a  3D relative velocity
of
$V=-1978$\,km\,s$^{-1}$ between the clusters and a 3D distance of
$R=1.31$\,Mpc.
Our FLASH simulations cannot constrain the relative velocity (or the
infall velocity) alone, but they indicate a pre-core passage state for
the cluster.

From these data it appears that the NE and 
SW components of Abell 2465  resemble the nearer and somewhat less
massive double cluster Abell 3716 (PLCKG345.40-39.34) studied by
Andrade-Santos \etal (2015), who conclude that that cluster is in a
pre-collisional state, and the binary clusters Abell 1750 and 1758
(Okabe \& Umetsu 2008) which  also appear to be in the early merger
stages. Consequently Abell 2465 is an excellent candidate for studying
the early stages of galaxy cluster collisions.

The physical separation of the galaxy, gas, and dark matter components is
of considerable current interest and has been modelled by many investigators
(\eg Schaller \etal 2015; Massey \etal 2015; Kim, Peter \& Wittman 2016).
For Abell 2465 this offset is
small and appears to be below the reliability of our data so we can only
place upper limits. The small 0.2\,Mpc weak-lensing to X-ray offset in 
Section~3.4, however would be comparable to that found in  some other clusters, 
except in the pre-merger state, one expects the gas
to trail the galaxies and dark matter. Simulations
of merging halos, (\eg by Vijayaraghavan
\& Ricker 2015) of merging halos show the stripped gas trails and galaxy
wakes which would follow rather than precede the sub-clusters.

It is interesting to note that although the weak-lensing and optical data
assign the higher mass to the NE sub-component, from weak lensing 
it has a more symmetrical
profile of mass contours. The SW sub-component with the more compact
central region has more unsymetrical and extended outer mass contours that 
show a flattened shape perpendicular to the line joining the two clumps. This
might be a sign of the merging of the two components. Since the 
subclusters are separated by
about 1.2\,Mpc, there should be substantial overlap of their
halos and therefore the possibility for interaction between the two
components. In their current configuration, the components of Abell 2465
appear to show no marked distortions or offsets.

\section{Conclusions}

The goal of this investigation was to determine the dynamical state of the
components of the double galaxy cluster Abell 2465. In advanced 
mergers these are known to separate. The X-ray data reveal
the distribution of the baryonic gas component and the weak lensing shows
the distribution of the total mass including dark matter. This can be
compared with the distributions of the luminous matter from given in
Papers I and II. From this study we draw the following conclusions.

(1) Using the \Chandra X-ray observations in Section~2,
the X-ray surface profiles of the NE and SW components were fit
employing modified $\beta$-models. The NE sub-cluster has a broader
profile than the SW component. 
The \Chandra X-ray data provided temperatures, gas and total masses, 
and gas fractions,
and determined a central entropy.  
This indicates that the NE and SW subclusters
have
gas masses 1.90 and $0.96\times 10^{13}M_\odot$ and
total masses 1.85 and $2.17 \times 10^{14}M_\odot$.  

(2) The entropy profiles from the X-ray data differ for the two subclumps. 
The NE central entropy is 
higher compared to the SW. This could be due to NE having undergone a recent 
merger. This seems consistent with NE having the three BCGs which should
merge in a short time and SW being more relaxed as it has a sharper central
profile and suggestive of a cool core.

(3) Section 2.5 shows that there are no large amounts of X-ray gas between
the two sub-clusters of Abell 2465, such as are found in colliding
clusters \eg the Bullet Cluster. 

(4) The weak-lensing analysis in Section~3 utilizes Subaru Suprime-Cam
$Vi'z'$ images for the background selection and shape measurements. A 
two-dimensional shear fitting with simultaneous modeling of the two
components of Abell 2465 was conducted. This confirms 
the corresponding weak-lensing virial masses are
3.7 and $2.9\times 10^{14}M_\odot$ and from redshift data are
4 and $3\times 10^{14}M_\odot$ (Paper I).
The projected mass contours are given in Figure~\ref{fig:map} 
which shows some distortion of the contours but only small offsets
between them and the X-ray gas. 

(5) From the optical redshift measurements in Paper I, the two
sub-clusters of Abell 2465 should be gravitationally bound.
The cluster as a whole is of an intermediate mass ($M \simlt 10^{15}M_\odot$).
The absence of strong X-ray emission between the two sub-components 
and no large offsets between the galaxies and weak lensing centers, 
suggest that the two subclusters have not yet collided (i.e pre-core
passage). Using the FLASH simulations and radial infall model indicates
that they will meet in $\sim 0.4$\,Gyr. 

It seems clear that the NE and SW components of Abell 2465 are in a 
pre-merger state. Although there is no large separation of the
collisionless and baryonic
components, such a system might provide constraints on their interactions 
as they begin to merge. Future observations that could better determine the
orbital parameters would be useful for detailed dynamical modeling.

\acknowledgements
The \Chandra observations were supported by NASA grant GO2-13150A to GAW,
who wishes to acknowledge correspondance on Abell 2465 from Ryan Johnson, 
Catherine Heymans, Eric Tittley, Meghan Gray, David Helfand, Robert Becker,
Adi Zitrin, Emilio Falco, Ho Seong Hwang, and William Dawson.
KU acknowledges support by the Ministry of Science and Technology of
Taiwan through grants MOST 103-2112-M-001-030-MY3 and MOST
103-2112-M-001-003-MY3.
WF and CJ acknowledge support from the Smithsonian Astrophysical Observatory
and contract NAS8-03060 from NASA.  
MN acknowledges support from PRIN-INAF 2014 1.05.01.94.02.


{}

\end{document}